\documentclass{aa}

\usepackage{graphicx}
\usepackage{xcolor}
\usepackage[colorlinks=true, linkcolor=blue, citecolor=blue, urlcolor=blue]{hyperref}
\usepackage{txfonts}

\newcommand{\rf}{$R_{\mathrm{final}}$}
\newcommand{\rb}{$R_{\mathrm{birth}}$}
\newcommand{\dsfr}{$\Delta\mathrm{SFR} = \mathrm{SFR}(R_{\mathrm{birth}}) - \mathrm{SFR}(R_{\mathrm{final}})$}
\newcommand{\dsfrf}{$\Delta\mathrm{SFR}_{\mathrm{fr}} = \Delta\mathrm{SFR} / \mathrm{SFR}(R_{\mathrm{birth}})$}

\begin{document} 

\title{The impact of radial migration on disk galaxy star formation histories}
\subtitle{I. Biases in spatially resolved estimates}

    \author{I.~Minchev \inst{\ref{aip}}, 
            K.~Attard \inst{\ref{sur}},
            B.~Ratcliffe \inst{\ref{aip}},
            M.~Martig \inst{\ref{liv}},
            J.~Walcher \inst{\ref{aip}}, 
            S.~Khoperskov \inst{\ref{aip}},
            J.P.~Bernaldez \inst{\ref{aip}},
            L.~Marques \inst{\ref{aip}},
            K.~Sysoliatina \inst{\ref{aip}},
            C.~Chiappini \inst{\ref{aip}},
            M.~Steinmetz \inst{\ref{aip}},
            R.S.~de Jong \inst{\ref{aip}}
            }

\institute{Leibniz-Institut f\"ur Astrophysik Potsdam (AIP), An der Sternwarte 16, D-14482, Potsdam, Germany\label{aip} \\
    \email{iminchev@aip.de}
     \and Astrophysics Research Group, University of Surrey, Guildford, Surrey GU2 7XH, UK \label{sur}
     \and Astrophysics Research Institute, Liverpool John Moores University, 146 Brownlow Hill, Liverpool L3 5RF, UK\label{liv}
    }
 
\abstract{Knowledge of the spatially resolved star formation history (SFH) of disk galaxies provides crucial insight into disk assembly, quenching, and chemical evolution. However, most reconstructions, both for the Milky Way and for external galaxies, implicitly assume that stars formed at their present-day radii. Using a range of zoom-in cosmological simulations, we show that stellar radial migration introduces strong and systematic biases in such SFH estimates, and in a Milky Way–like case study we link these biases directly to the disk’s merger-driven, non-axisymmetric response. In the inner disk ($R \lesssim h_d$), early star formation is typically underestimated by 25–50\% and late star formation overestimated, giving the misleading impression of prolonged, moderate activity. An exception occurs in the most central bin considered ($\sim0.4h_d$), which is consistently overestimated due to a net inflow of inward migrators. At intermediate radii and in the outer disk, migration drives the opposite trend: intermediate-age populations are overestimated by 100–200\% as stars born in the inner disk migrate outward, whereas genuinely in situ populations are underestimated by $\sim50\%$ as they themselves continue to migrate. The net effect is that SFH peaks are suppressed and broadened, and the true rate of inside-out disk growth is systematically underestimated. These distortions affect all galaxies in our sample and have direct implications for interpreting spatially resolved SFHs from integral field unit surveys such as CALIFA and MaNGA, where present-day radii are often used as proxies for stellar birth sites. Correcting these biases will require accounting for the disk mass, bar presence, disk kinematics and morphology, and recent birth-radius estimation techniques for Milky Way stars offer a promising path forward.
} 


\authorrunning{Minchev, Attard, et al.}
\titlerunning{Impact of radial migration on disk galaxy SFHs}
\maketitle

\section{Introduction} 

Estimates of the spatially resolved star formation history (SFH) of disk galaxies, including the Milky Way, are central to understanding galaxy formation and evolution. In nearby systems, stellar age distributions have typically been derived through color–magnitude diagram (CMD) modeling, using apparent luminosities and distance estimates \citep[e.g.,][]{gallart99,aparicio09}. This approach has been applied to galaxies such as M31 \citep{bernard15}, the Magellanic Clouds \citep{weisz13,cohen24}, and NGC~7792 \citep{sacchi19}, and in the Milky Way to the bulge \citep{bernard18} and local disk \citep{ruiz-lara20,gallart24,fernandez25, delalc25} with \emph{Gaia} data \citep{gaia18,drimmel23}. These CMD-based results complement other SFH reconstruction techniques, such as fitting the observed age–[$\alpha$/Fe] relation with chemical evolution models \citep[e.g.,][]{chiappini97,snaith15,haywood16}. Alternatively, the star formation rate (SFR) can be derived within a more comprehensive joint Bayesian inference framework where CMD information is an integral part of constraining the SFR and other Milky Way model parameters, alongside spatial and kinematic data \citep{sysoliatina21}. 

For galaxies beyond the Local Group, the standard method for recovering a resolved SFH is spectral fitting of integrated light. Earlier work was reviewed by \citet{walcher11}. More recent studies using large integral-field surveys such as the Calar Alto Legacy Integral Field Area survey (CALIFA) and Mapping Nearby Galaxies at Apache Point Observatory (MaNGA) have determined the spatially resolved SFHs for thousands of galaxies \citep[e.g.,][]{cid13,gonzalez14,gonzalez17, wilkinson15, sanchez19,peterken20}. These results generally support the inside-out growth scenario \citep{perez13} and the persistence of the star formation main sequence to early cosmic times \citep{sanchez19}.  

In both external galaxies and the Milky Way, the ultimate goal is to recover the distribution of stellar ages as a function of galactocentric radius, SFH($R$). After correcting for survey selection effects and observational biases and accounting for the lifetimes of massive stars via the initial mass function, such estimates can be linked to the disk growth rate and the radial dependence of gas accretion and chemical enrichment \citep{matteucci89,prantzos95, chiappini97,frankel19}. Bursts in the SFH may further trace disk–satellite interactions or episodes of gas inflow \citep{SotilloRamos22,dicintio21, khoperskov23, annem24, zibetti24,wang24}, as well as instabilities occurring during bar buckling     \citep{nepal24}.

\begin{figure*}
\centering
        \includegraphics[width=1.9\columnwidth]{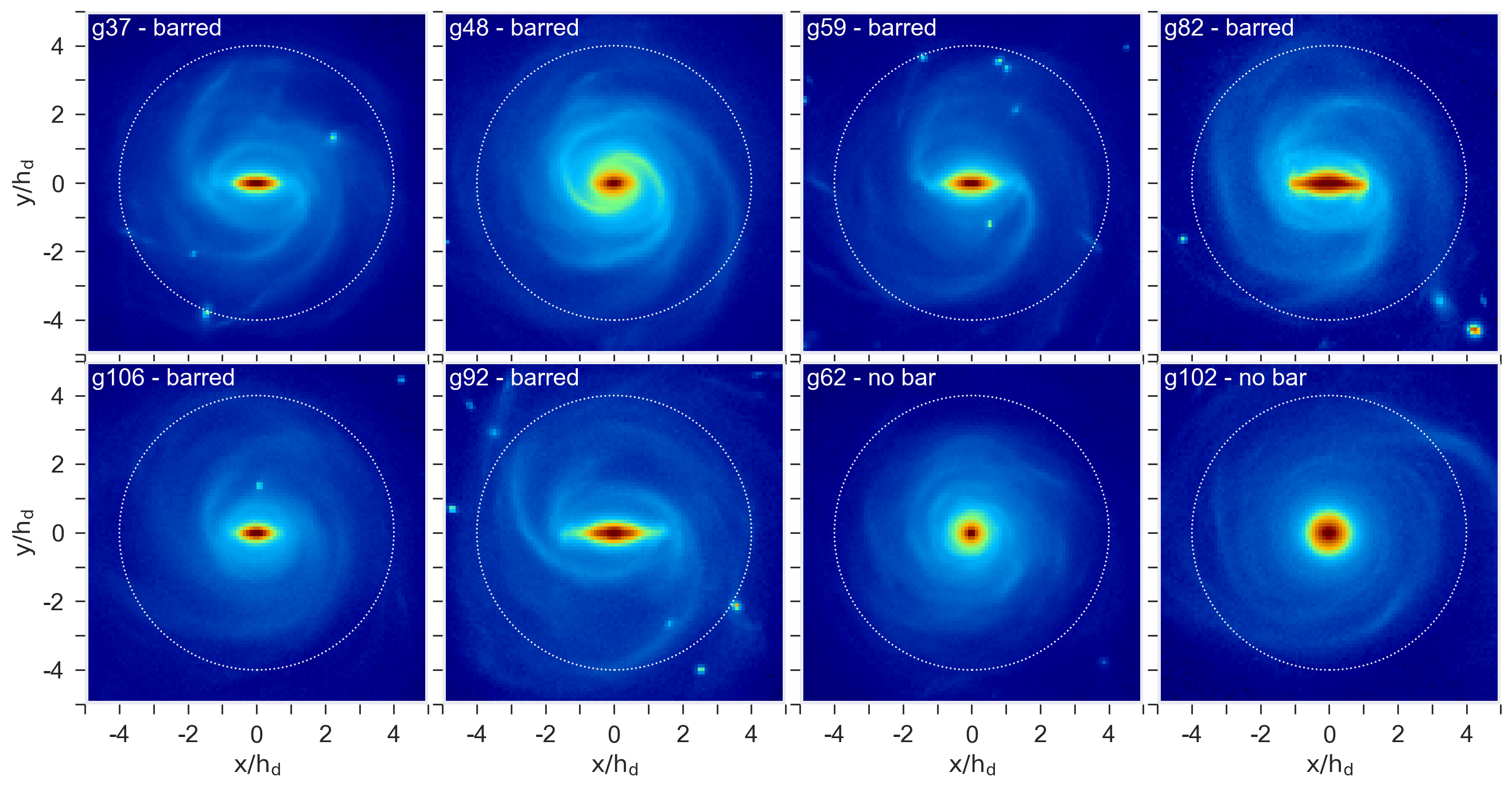}
    \caption{Face-on views of all galactic disks at the final simulation time, in units of scale lengths, $h_d$. The dotted circles indicate the outermost radial bin ($4h_d$) considered in our analyses.
    }
    \label{fig:faceon}
\end{figure*}

A critical assumption underlying these efforts is that stars remain close to their birth radii, $R_{\mathrm{birth}}$. In reality, galactic disks host bars, spiral arms, and satellites that redistribute stellar angular momentum and drive radial migration \citep{sellwood02, roskar08a, minchev11a,vera-ciro14, khoperskov20, marques25, zhang25}. For example, \citet{baba25} used Milky Way–focused simulations to show that stellar age distributions at a fixed radius can depart strongly from the in situ SFH, due to the outward migration of stars born in the inner disk rather than a local SFR burst. Observations now indicate that a majority of stars in the solar vicinity were born at different radii and subsequently migrated, rather than having formed locally (e.g., \citealt{minchev18, frankel20,lu24, dantas25}).

This process, also known as radial mixing, has been predicted in simulations for decades \citep[e.g.,][]{wielen77, friedli93,debattista06,mq06} and is now recognized as essential for explaining chemo-kinematic observations in the Galaxy: the scatter and apparent flatness of the age–metallicity relation \citep{schonrich09a,mcm13,kubryk15b, khoperskov20, prantzos23}, the metallicity distribution of the solar neighborhood \citep{hayden15,loebman16,miglio21}, the flattening of radial abundance gradients with stellar age \citep{anders17b,renaud25,ratcliffe25}, and the structure of the [$\alpha$/Fe]–[Fe/H] plane \citep{queiroz20, ratcliffe23, ratcliffe25, khoperskov2025_2}.  

In addition to large-scale processes, stellar scattering off smaller structures can also redistribute orbits and shape galactic disks. Cloud--star scattering can generate and maintain exponential stellar profiles \citep{wu20}, while disk-halo gas flows \citep{struck18} and interstellar holes and clumps \citep{struck17} further redistribute stars, particularly in dwarf irregulars where star formation appears to occur from the outside in \citep{zhang12}. Stochastic scattering models show that exponential profiles are a robust outcome of generic disk-like scattering \citep{elmegreen16}. These processes are especially important in the outer regions of disks, where spirals and bars are less effective, highlighting the combined role of large- and small-scale mechanisms in shaping galactic structure.

In this study we quantified how stellar radial migration affects the recovery of spatially resolved SFHs in galactic disks. Using a set of eight cosmological simulations with a range of morphologies and evolutionary histories, we traced stars from their true birth radii to their present-day positions. This allowed us to assess the magnitude, radial dependence, and systematic nature of the migration-induced biases, and to explore their implications for both the Milky Way and external integral field unit (IFU) surveys.  

\begin{figure*}
\centering
        \includegraphics[width=1.9\columnwidth]{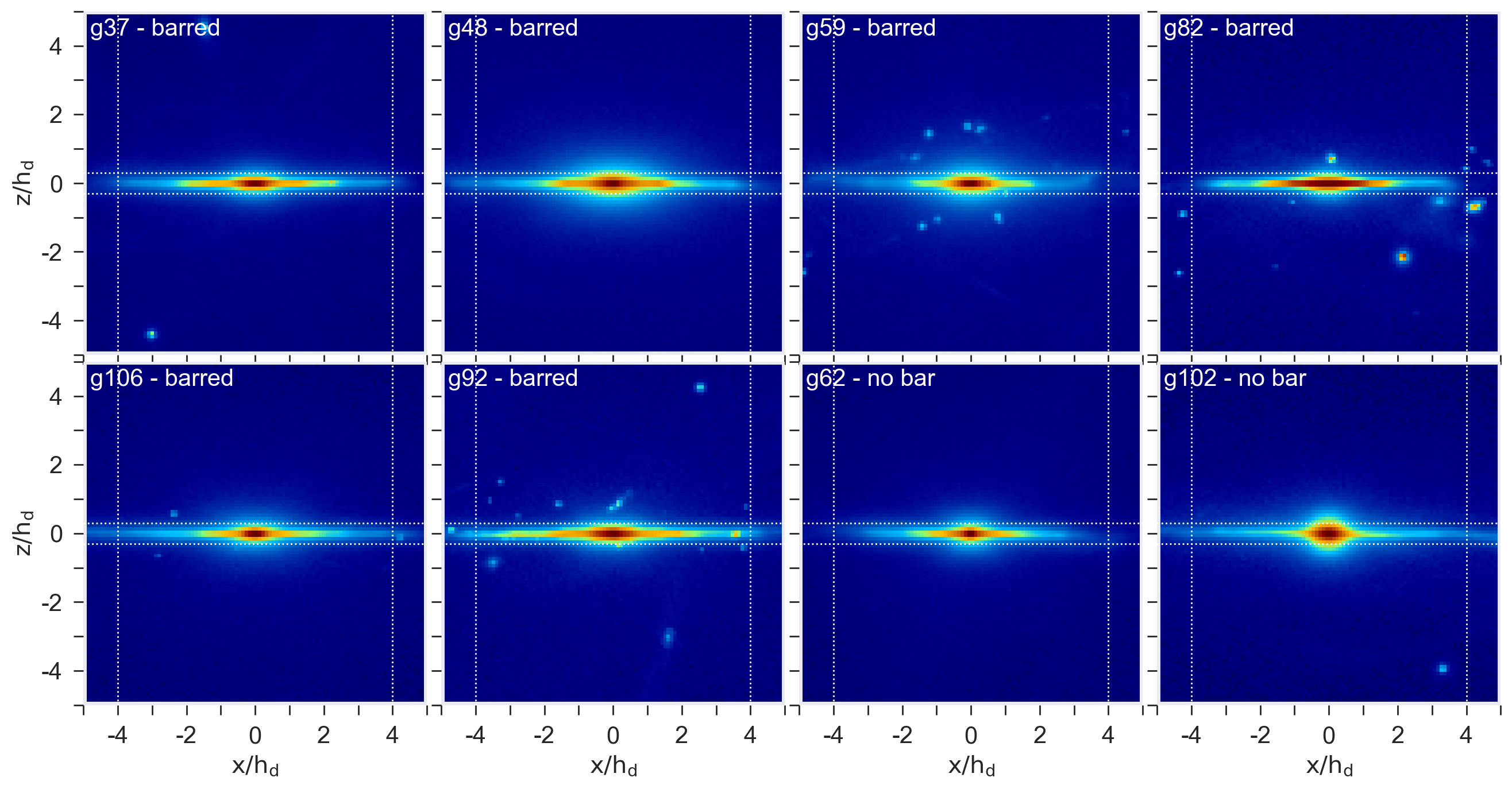}
    \caption{Edge-on views of all galactic disks at the final simulation snapshot, shown in units of disk scale length, $h_d$, as in Fig.~\ref{fig:faceon}. The dotted vertical lines mark the outermost radial bin ($4h_d$) included in our analysis. The horizontal lines indicate the vertical birth cut, $|z_0| \leq 0.3h_d$, used for all SFH estimates, which ensures that stars are selected as being born in the disk.
    }
    \label{fig:edgeon}
\end{figure*}

\begin{table*}[htbp]
\caption{Properties of the simulations used in this work.}
\label{table}
\centering
\begin{tabular}{lccccc}
\hline\hline
Name & $h_d$ [kpc] & Bar presence & Mass [$10^{10} M_{\odot}$] & B/T & 
Merger events (look-back time [Gyr], mass ratio) \\
\hline
g37  & 7.5 & yes & 12.0 & 0.13 & No massive merger \\
g48  & 4.5 & yes & 10.8 & 0.07 & 5.1 Gyr, 1:5 \\
g59  & 5.0 & yes & 7.1  & 0.28 & 6.2 Gyr, 1:4; 7.8 Gyr, 1:12; 8.0 Gyr, 1:2 \\
g62  & 4.8 & no  & 6.6  & 0.16 & 8.8 Gyr, 1:10; 9.0 Gyr, 1:1 \\
g82  & 5.5 & yes & 3.8  & 0.02 & No massive merger \\
g92  & 5.5 & yes & 4.3  & 0.04 & No massive merger \\
g102 & 4.0 & no  & 3.3  & 0.48 & 6.0 Gyr, 1:21 \\
g106 & 5.1 & yes & 4.3  & 0.22 & 7.1 Gyr, 1:14 \\
\hline
\end{tabular}

\tablefoot{
Listed are the disk scale length,  $h_d$, bar presence, total stellar mass, bulge-to-total ratio (B/T), and the merger history for events with mass ratios higher  than 1:50 occurring within the last 9 Gyr (look-back time).
}
\end{table*}

\section{Simulations}
\label{sec:sims}

The simulations we studied were presented by \citet{martig09,martig12}. They come from a suite of 33 galaxies simulated from redshift 5 with redshift zero stellar masses ranging from $1\times 10^{10}$ to $2\times 10^{11}$\,M$_{\sun}$. Dark matter haloes are selected in a large volume, dark matter-only simulation, performed using $\Lambda$-cold dark matter ($\Lambda$CDM) cosmology with the adaptive mesh refinement code RAMSES \citep{teyssier02}. The boundary conditions used for the zoom-in simulation replicate all minor and major mergers as well as diffuse infall, as imposed by the initial cosmological simulation. 

These zoom-in simulations use a spatial resolution of 150~pc and a mass resolution of $10^{4-5}$\,M$_{\sun}$ ($1.5\times10^4$\,M$_{\sun}$ for gas, $7.5\times10^4$\,M$_{\sun}$ for stars and $3\times10^5$\,M$_{\sun}$ for dark matter particles). Stars formed during the simulation have the same mass as gas particles. To model gravity, the particle-mesh code described in \cite{BC02} was used. Star formation is computed with a Schmidt-Kennicutt law \citep{kennicutt98} with an exponent of $1.5$ and an efficiency of $2\%$. The star formation threshold is set at $0.03$\,M$_{\sun}$pc$^{-3}$. Energy feedback from supernovae explosions using a kinetic scheme as well as the continuous gas mass-loss from stars \citep{martig10} are included. More details on the zoom-in simulation technique can be found in \cite{martig09, martig12}.

For this study we selected eight galaxies that span a range of bar sizes (or lack of a bar), disk scale lengths and masses. We refer to different models by their simulation number (g48, g59, etc.) We worked in units of disk scale lengths, $h_d$, to facilitate comparisons between the different simulations. The masses and scale lengths are listed in Table~\ref{table}, as measured by \cite{martig12}. The models chosen for this work include five barred and three non-barred galaxies with face-on and edge-on morphologies shown in Figs.~\ref{fig:faceon} and \ref{fig:edgeon}, respectively. Central bars are seen in g37, g48, g59, g82, g106, and g92 but not in g62 or g102. The strongest bars in terms of disk scale length is clearly seen for g92 ($\sim1.8h_d$), followed by g82. Many of the properties of these bars, such as their formation and evolution, are explored by \cite{kraljic12}, and their effect on the disk dynamics has been studied quite extensively, for example the radial migration in g92 (\citealt{minchev12b}), disk heating using the full suite (\citealt{martig14a, martig14b}), and the effect of bar fluctuations on velocity field, radial migration, and bar fluctuations in g106  (\citealt{carrillo18,hilmi20, vislosky24,marques25}).

At the last snapshots shown in Fig.~\ref{fig:faceon} all galaxies show the presence of spiral arms, mostly dominated by $m=2$ modes, except for g62 and g102 where multiple structures are seen. Possible matches to the Milky Way in terms of morphology would be g37, g106, and g92.

All galaxies in our sample experience significant merger activity at early times, consistent with expectations from $\Lambda$CDM cosmology and results from cosmological simulations. In particular, prior to 9 Gyr look-back time, all systems undergo multiple mergers, reflecting the hierarchical growth characteristic of galaxy formation. Satellites remain visible in the last snapshot shown in Fig.~\ref{fig:faceon} and especially in Fig.~\ref{fig:edgeon}. Table \ref{table} summarizes the merger histories, listing only events with mass ratios more significant than 1:50 and occurring within the last 9 Gyr (look-back time). While some galaxies (e.g., g37, g82, and g92) are merger-free in this epoch, others such as g59 and g62 experience multiple interactions, including mergers as significant as 1:2 and 1:1. We expect such events -- whether minor or major -- to impact the SFH both by gravitationally perturbing the gas disk and by delivering fresh gas to the host galaxy, in addition to contributions from filamentary inflows and flyby interactions.

\begin{figure*}
\centering
        \includegraphics[width=1.9\columnwidth]{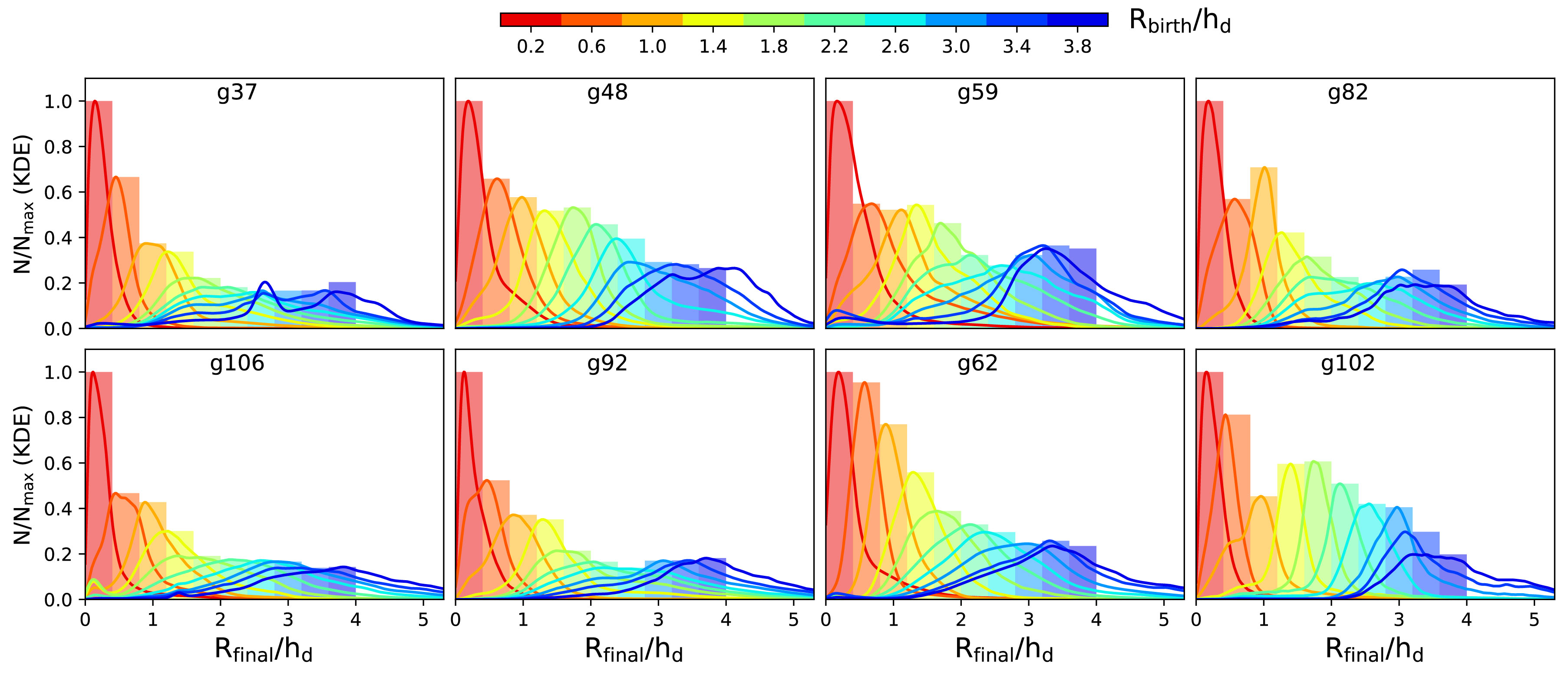}
    \caption{Stellar final-radius distributions for ten birth radial bins of width $0.4h_d$, where $h_d$ is the radial scale length of each galaxy measured at the final time (Table~\ref{table}). Significant radial migration is present in all disks. The \rf\ distributions are skewed toward larger radii for inner-disk bins and toward smaller radii for outer-disk bins. Overall, the disks extend by roughly one additional scale length owing to angular-momentum transport.
    }
    \label{fig:rbirth}
\end{figure*}

\section{Results}
\label{sec:results} 

\subsection{Radial migration in the simulations}

To quantify the extent of radial migration over each galaxy's lifetime, we started by dividing the stellar disk into ten equally spaced birth-radius annuli of width $0.4h_d$, thereby covering four disk scale lengths in total. To ensure that we selected stars genuinely formed in the disk, we restricted the sample to those with vertical birth positions within $|z_0| \leq 0.3h_d$, where $z_0$ is the distance from the mid-plane at the time of formation.

In Fig.~\ref{fig:rbirth}, each panel shows the distribution of final radii, \rf, for stars born in a given annulus and the corresponding birth radius, \rb. The maximum of each \rf\ distribution determines the height of the \rb\ bar. Across all models, the distributions remain relatively symmetric at intermediate radii ($\sim 2h_d$), but become increasingly skewed in both the inner and outer disk. Stars with \rb$< 1.5h_d$ tend to migrate outward, while those formed in the outer disk (\rb$> 2.6h_d$) develop extended two-sided tails, often with peaks shifted inward. Notably, the disks extend by more than a scale length primarily due to stars born outside $2.6h_d$.

These trends demonstrate that radial migration is a major factor in shaping the present-day structure of all eight disks. While most \rf\ distributions still peak close to the stars' birth radii, a significant fraction of stars have migrated, as seen in the broad and asymmetric tails. Both the magnitude and the direction of migration vary systematically with radius, implying that spatially resolved SFHs inferred at the present time will be biased unless migration effects are accounted for \citep{roskar08b, brunetti11, minchev12a}. Crucially, such corrections must be radially dependent to capture the nonuniform impact of migration across the disk, as illustrated by \cite{mcm14}.

\subsection{SFHs determined using \rb\ versus \rf}
\label{sec:sfh}

Using the same radial bins as in Fig.~\ref{fig:rbirth}, we computed the SFR surface density, defined as the mass of stars formed per unit disk area per unit time, with units of $M_{\odot}\,\mathrm{pc}^{-2}\,\mathrm{Gyr}^{-1}$.

\begin{figure*}
    \centering
        \includegraphics[width=1.6\columnwidth]{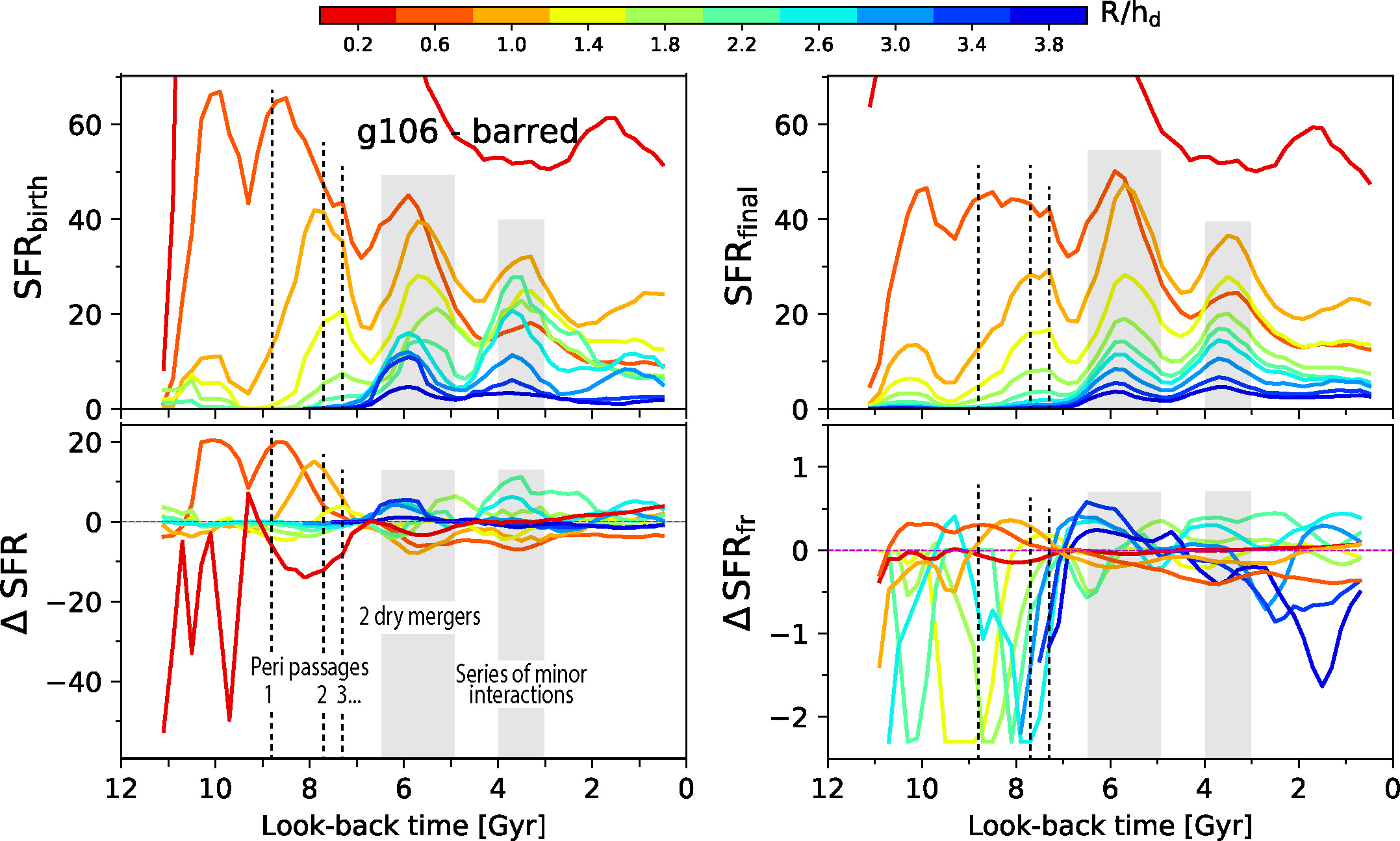}
    \caption{
{\bf Top:} SFR for galaxy simulation g106, computed in bins of birth radius (left) and final radius (right), using the ten radial bins defined in Fig.~\ref{fig:rbirth}. SFRs are measured in 1~Gyr-wide age bins, spaced by 0.2~Gyr, and have units of $M_{\odot}\,\mathrm{pc}^{-2}\,\mathrm{Gyr}^{-1}$. To select stars formed in the disk, for all plots we apply a vertical cut at birth, $|z_0| \leq 0.3h_d$, where $h_d$ is the disk scale length. To improve visibility across bins, the vertical axis is normalized to the second innermost bin. {\bf Bottom:} Absolute difference between the SFH computed from birth and final radii, \dsfr, shown on the left, and the corresponding fractional difference, \dsfrf, shown on the right. The largest absolute discrepancies occur in the inner disk at early times (red and orange curves, bottom left), whereas the fractional differences can exceed 200\% in the mid and outer disk. The vertical dotted lines mark the first three pericentric passages of the last massive gas-rich merger. The shaded regions indicate the disk response to later dry mergers and a phase of ongoing minor interactions. These external events coincide with pronounced features in $\Delta \mathrm{SFR}$ and $\Delta \mathrm{SFR}_{\mathrm{fr}}$, highlighting their role in triggering both enhanced star formation and accelerated stellar migration (see Fig.~\ref{fig:fourier} for the corresponding dynamical signatures).
}
    \label{fig:sfh1}
\end{figure*}

 The top-left panel of Fig.~\ref{fig:sfh1} shows the SFR as a function of look-back time for model g106, evaluated at the stellar birth radius. We refer to this as the true SFH, or $\mathrm{SFR}(R_{\mathrm{birth}})$. For stars born at \rb$\lesssim 1h_d$ (red curves), the SFR peaks between 7 and 10 Gyr ago. At larger birth radii, star formation occurs progressively later, consistent with an inside-out disk growth scenario \citep{matteucci89, chiappini97, mcm13, frankel19, prantzos23}.

The first prominent SFR peak for stars with median \rb$\lesssim 1h_d$ (red curves) in g106 arises from the initial coalescence of two massive, gas-rich satellites with the protogalaxy. The most significant merger after look-back time $lbt = 9$~Gyr begins with a pericentric passage at $lbt \sim 8.8$ Gyr, involving a satellite with a mass ratio of 1:14 (see Table\ref{table}). This encounter occurs within the inner scale length (measured at the final time), proceeds along a polar orbit, and can be linked to the second SFR peak in the top-left panel of Fig.\ref{fig:sfh1}. The second pericentric passage follows at $lbt \sim 7.7$~Gyr, during which the orbit is strongly deflected by the central potential, transitioning from polar to in-plane -- a behavior commonly seen in simulations. This allows the merger to affect the disk at radii outside $1h_d$. A third passage occurs at $lbt\sim 6.4$~Gyr, after which the satellite is rapidly disrupted by $lbt\sim 6.7$~Gyr. These three pericentric passages (marked by the vertical dotted lines in Fig.~\ref{fig:sfh1}) coincide with the SF peaks in the birth-radius bins centered at $0.6 h_d$, $1.0 h_d$, and $1.4 h_d$.

\begin{figure*}
        \centering
    \includegraphics[width=0.9\textwidth]{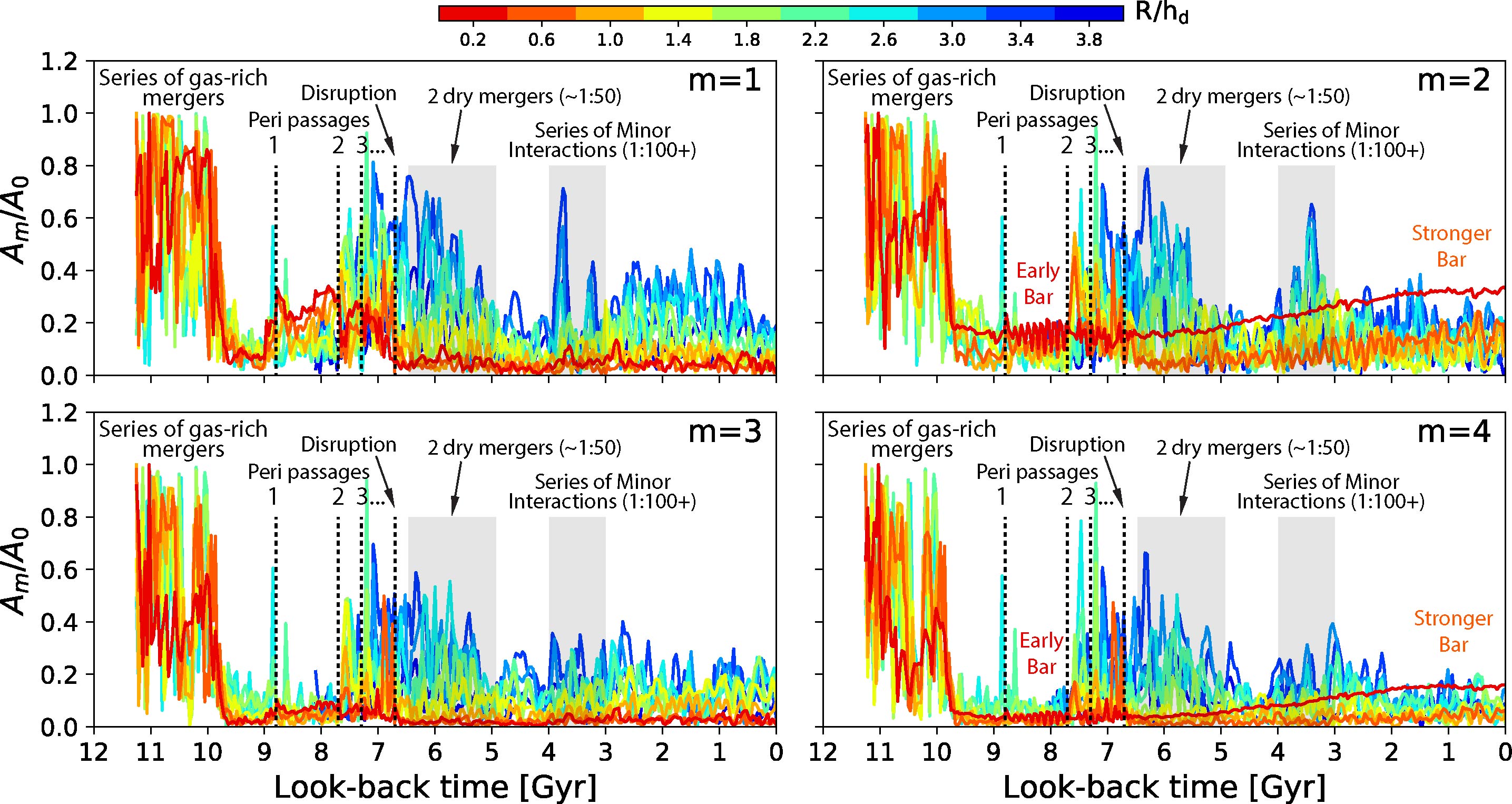}
    \caption{
Time evolution of normalized Fourier amplitudes, $A_m / A_0$, for $m = 1$ to $m = 4$ in simulation g106. Vertical dotted lines mark the first three pericentric passages of the last massive gas-rich merger, followed by a rapid sequence of additional close passages and complete disruption by $lbt \sim 6.7$~Gyr. The shaded region to the left in each panel highlights the influence of two dry mergers (mass ratio of $\sim$1{:}50), with pericenters between $lbt \sim 6.5$ and $4.9$~Gyr. The right shaded region indicates a more chaotic phase, driven by a sequence of minor interactions (mass ratio $\gtrsim$1{:}100). Bursts of multi-mode activity are seen near these events. Radial bins are color-coded as in Fig.~\ref{fig:sfh1}. The timing of mode amplification closely mirrors SFR peaks, suggesting a causal link between external perturbations and disk structure.
    }
    \label{fig:fourier}
\end{figure*}

After the disruption of this last massive gas-rich satellite, g106 simultaneously undergoes two further, predominantly dry mergers with mass ratios of order $1{:}50$, whose pericenters lie between $lbt \sim 6.5$ and 4.9~Gyr (left shaded regions in Figs.~\ref{fig:sfh1} and \ref{fig:fourier}). Because these satellites bring relatively little fresh gas, their imprint on ${\rm SFR}(R_{\rm birth})$ is more modest than that of the $1{:}14$ event. Nevertheless, they prolong the period of elevated star formation in the inner and intermediate bins and re-excite non-axisymmetric structure throughout the disk, producing secondary undulations that are visible in both ${\rm SFR}(R_{\rm birth})$ and ${\rm SFR}(R_{\rm final})$.

Finally, the last prominent peak in the true SFH of g106 (top-left panel of Fig.~\ref{fig:sfh1}), at $lbt \sim 3.5$~Gyr, exhibits a stronger response in the intermediate and outer radial bins (blue and cyan curves), also suggesting an external trigger. Although a single, well-defined interaction is not apparent, the movie of this galaxy reveals numerous minor satellites in the vicinity at that time, which collectively perturbed the disk. Figure~\ref{fig:fourier} shows a clear amplification of $m=1$ and $m=2$ modes at this time, indicating a collective disk response that we examine in more detail in Sect.~\ref{sec:modes_discussion}. This behavior illustrates how even weak external perturbations can imprint distinct, localized fluctuations on top of the otherwise smooth inside-out growth of the disk.

\begin{figure*}
        \centering
    \includegraphics[width=\textwidth]{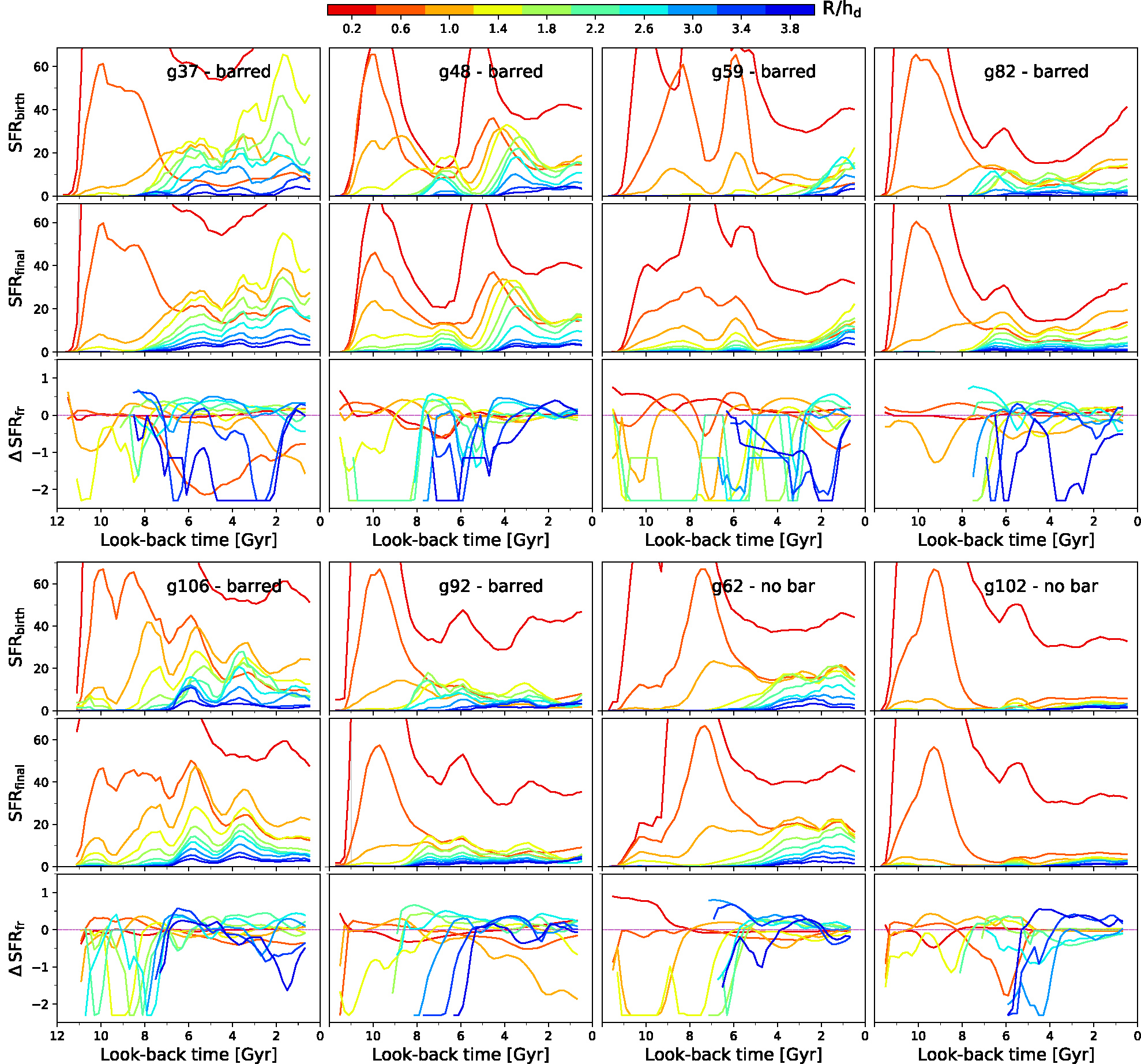}
    \caption{
SFHs for the full galaxy sample. 
Each system is shown in a vertical three-panel block. 
{\bf Top:} $\mathrm{SFR}(R_{\mathrm{birth}})$, the SFH binned by stellar birth radius. 
{\bf Middle:} $\mathrm{SFR}(R_{\mathrm{final}})$, the SFH binned by present-day stellar radius. 
{\bf Bottom:} Fractional difference between the two, $\Delta \mathrm{SFH}_{\mathrm{fr}}$. Across the sample, a consistent pattern emerges: early inner-disk star formation is underestimated in $\mathrm{SFR}(R_{\mathrm{final}})$, while outer-disk star formation is strongly overestimated at intermediate ages, with fractional biases often exceeding 200\%. These systematic effects reflect the redistribution of stars via radial migration.
    }
    \label{fig:sfh_all}
\end{figure*}

To examine how an observer would infer the spatially resolved SFH from the final stellar radii, the top-right panel of Fig.~\ref{fig:sfh1} shows the SFR as a function of look-back time, using the same binning scheme as in the left panel but sorted by present-day radius, $\mathrm{SFR}(R_{\mathrm{final}})$. Strong discrepancies are immediately apparent: the initial burst of star formation in the inner disk (red and orange curves) is both diminished in amplitude and shifted toward younger look-back times. A further striking feature is that distinct star-formation peaks, clearly separated in $\mathrm{SFR}(R_{\mathrm{birth}})$, appear aligned once radial mixing has redistributed stars, as in $\mathrm{SFR}(R_{\mathrm{final}})$. This behavior is consistently reproduced across our full sample of simulated galaxies (Fig.~\ref{fig:sfh_all}).

The bottom-left panel of Fig.~\ref{fig:sfh1} shows the absolute difference in SFH, defined as \dsfr, for each radial bin. The bottom-right panel displays the corresponding fractional difference, \dsfrf. A positive value of $\Delta \mathrm{SFR}_{\mathrm{fr}}$ indicates that using \rf\ underestimates the true SFR, and vice versa. In estimating $\Delta \mathrm{SFR}_{\mathrm{fr}}$, we set a threshold to avoid excessive values. This can be seen in the square appearance at $\Delta \mathrm{SFR}_{\mathrm{fr}}=-2.3$ for all galaxies (Fig.~\ref{fig:sfh_all}), where each system is shown in a three-panel block comparing $\mathrm{SFR}(R_{\mathrm{birth}})$, $\mathrm{SFR}(R_{\mathrm{final}})$, and $\Delta \mathrm{SFH}_{\mathrm{fr}}$.

As expected, the largest discrepancies in stellar mass (bottom-left panel of Fig.~\ref{fig:sfh1}) arise in the innermost radial bins, where stellar densities and SFRs are highest and where large-scale migration driven by mergers is most frequent at high redshift. The strong exchange between the inner two \rb\ bins is apparent both from the near-mirroring of their curves across the $\Delta \mathrm{SFR}=0$ line and from their overlapping \rf\ distributions in Fig.~\ref{fig:rbirth}. Stars born at $h_d=1.4$-$2.6$ migrate inward, giving the false impression of a more extended and less intense SFH in the inner disk. Consequently, the oldest bimodal peak in the true SFH is diminished by $\sim 20\,M_{\odot}\,\mathrm{pc}^{-2}\,\mathrm{Gyr}^{-1}$, corresponding to a relative difference of about 30\%, as shown in the bottom-right panel of Fig.~\ref{fig:sfh1}.

Further examining $\Delta \mathrm{SFH}_{\mathrm{fr}}$ (bottom-right panel of Fig.\ref{fig:sfh1}), we find systematic biases across the disk. At intermediate ages ($lbt \sim 7$-11~Gyr), the outer regions show strong overestimates of the SFR, reaching 100-200\%, while locally born stars in these bins are underestimated by about 50\%. At later times ($lbt \lesssim 6$~Gyr), the intermediate bins become underestimated, while the apparent SFR in the outskirts continues to rise, reaching $\sim$100\% at $lbt \sim 1.5$~Gyr.

A strong exchange is also visible across $R \sim 1$-$1.4\,h_d$, where inner-disk curves (red) are underestimated by $\sim$30\% while intermediate curves (orange-cyan) are simultaneously overestimated by up to 200\%. This indicates substantial redistribution of stellar mass across this radius. Similar patterns are found in the other galaxies of our sample (Fig.~\ref{fig:sfh_all}), with the common feature that the outer disk SFR is consistently overestimated.

\section{Discussion}

\subsection{Dynamical response to mergers in g106}
\label{sec:modes_discussion}

To further illustrate the link between the SFH and the dynamical evolution of the disk, we analyzed the time evolution of non-axisymmetric Fourier modes in g106. Figure~\ref{fig:fourier} shows the amplitudes of the $m = 1$ to $m = 4$ modes, normalized by the axisymmetric component $A_0$, as a function of look-back time. 

The coalescence of two massive, gas-rich mergers that trigger initial disk formation is seen as a strong disturbance in all four modes at $lbt \lesssim 9.5$~Gyr. The first three pericentric passages associated with the last, gas-rich, massive merger align closely with the prominent SFR peaks for the birth-radius bins centered at $h_d\approx 0.6$, $1.0$, and $1.4$, respectively, in Fig.~\ref{fig:sfh1}. 

These encounters excite a strong lopsided response ($m=1$), bar-like modes ($m=2,4$), and multi-armed spiral structure ($m=3,4$). The resulting patterns propagate from the inner to the outer disk as the initially polar satellite orbit gradually aligns with the disk plane by the third pericenter passage. As the amplitude of the outer spirals begins to decline around the time of the satellite’s full disruption, two smaller dry satellites start to interact with the host at intermediate and outer radii ($R_{\rm birth} > 1.5h_d$), with pericenters between $lbt \sim 6.5$ and $4.9$, as indicated by the gray strip in Fig.~\ref{fig:fourier}.

At later times ($lbt \sim 3.5$~Gyr), a sequence of minor flybys and dry mergers leads to a broad increase in amplitude across all modes, indicating that even weak perturbations can collectively induce measurable asymmetries in the stellar disk. Evidence that this is driven by external activity is that the $m=1$ peak (deviation from bisymmetry) precedes the rise in $m=2$ and $m=4$. The strong correlation with the offsets in $\Delta \mathrm{SFH}$ further indicates that these interactions not only enhance star formation but also drive significant radial migration -- the latter is suggested by the symmetric exchange of mass between the inner and outer radii seen in $\Delta \mathrm{SFH}$. 

It is intriguing that an early bar is triggered by the first pericentric passage of the last massive merger at $lbt \approx 8.8$~Gyr, visible as oscillatory behavior in $m=2$ and $m=4$ in the innermost bin (red curve). These high-frequency, periodic fluctuations in amplitude (as well as in bar length and pattern speed) are driven by bar–spiral interactions \citep{hilmi20, vislosky24}, which efficiently redistribute angular momentum at the bar–spiral interface \citep{mf10, marques25, kwak26a}. This initial bar is strongly perturbed by the subsequent passages and final disruption of the satellite (second and third peri in Fig.~\ref{fig:fourier}), after which a more stable phase of bar growth ensues, with oscillations shifting to larger radii (orange curves).

The early bar--spiral interaction, occurring in the very inner disk, is likely responsible for the nearly symmetric mass exchange observed at $lbt \approx 8$~Gyr in the $\Delta \mathrm{SFH}$ (bottom-left panel of Fig.~\ref{fig:sfh1}), where the innermost radial bins exhibit mirrored offsets about zero. This symmetry suggests an internal redistribution mechanism that is consistent with torque-driven angular momentum transfer. 

\subsection{Merger imprints on the SFH}

The wavy behavior of SFR with look-back time can be directly related to merger perturbations -- both through the gas they bring in (especially at high redshifts) and through their dynamical impact on the host disk, which can trigger star formation (e.g., \citealt{scannapieco09, nuza19,khoperskov23}). In g106, for example, we show that pericentric passages from the last massive merger could be linked to three strong peaks around $lbt=6-9$~Gyr, whereas the later peak at $lbt \sim 3.5$~Gyr reflects the cumulative effect of several minor satellite perturbations.

Late major mergers, as in the case of g48, produce pronounced steps in the SFR around the time of the merger, with an extreme example seen in the strongly perturbed g59. By contrast, the most quiescent galaxy in our sample, g92 (see \citealt{martig14a}), displays a smoother SFR evolution with only three mild peaks.

Large peaks are often systematically offset in time across radii. For example, the outer disk shows a delay of about 1 Gyr relative to the inner disk in g48 at $lbt \sim 4$~Gyr and in g106 at $lbt \sim 6$~Gyr. This behavior is likely caused by pericentric passages of minor mergers with specific orbital properties. In contrast, smaller satellites tend to induce more localized, randomly distributed peaks in the SFR at different radii, as seen in g82 at $lbt \sim 3-7$ Gyr. After migration, these peaks tend to align in time (second panel of each block in Fig.\ref{fig:sfh_all}).

Bars provide another pathway to drive late-time star formation in the inner disk, provided gas is available. In g82, for example, a strong bar develops only 1-2 Gyr before the end of the simulation, triggered by a flyby interaction. Although its present-day morphology is indistinguishable from that of a galaxy where the bar formed earlier (see Fig.~\ref{fig:faceon}), the bar's late emergence caused a sharp increase in SFR within the innermost radial bins. Although bars are typically found to form early in both simulations and observations, there have been indications of later bar formation in both external galaxies \citep{desafreitas25} and the Milky Way (e.g., \citealt{minchev10,nepal24}).

The structures described above suggest that variations in SFR with time and radius can be used to link disk SFHs to merger histories, including the number, timing, and orbital geometry of mergers (e.g., the angular momentum orientation, or whether the orbit is coplanar or polar). Such parameters are known to influence both stellar kinematics and star formation patterns \citep[e.g.,][]{Kazantzidis08, Moster10, Pontzen17, khoperskov19}. While we highlighted these connections in the case of g106, a systematic exploration of how merger parameters map onto SFH features across the entire sample will be the subject of future work.

\subsection{Implications for inside-out disk formation recovery}

We find that, in general, stellar migration tends to bias SFR estimates within one disk scale length, leading to underestimation at early cosmic times and overestimation at later times (Figs.~\ref{fig:sfh1}, \ref{fig:sfh_all}). This produces the misleading appearance of prolonged, moderate inner-galaxy star formation, as the genuine peak is dampened and temporally broadened. While the relative differences ($\Delta \mathrm{SFH}_{\mathrm{fr}}$) are typically smaller in the innermost bins than in the outskirts, the absolute impact on stellar mass is greatest in the inner disk owing to the steep exponential surface density profile of galactic disks. 

The SFR outside the inner disk scale length is systematically misestimated, with strong overestimates at old and intermediate ages and deficits at young ages. This arises naturally from the outward transport of angular momentum in disks \citep{lb72,minchev11a}, predominantly affecting the outer regions at early times and the inner regions at later times, in line with an inside-out growth scenario.

More specifically, the outer disks are strongly overestimated (100--200\%) at intermediate ages, when stars are not born there but migrate outward from the inner disk. By contrast, stars born locally at intermediate radii and the outskirts are typically underestimated by about 50\%, as they themselves migrate further outward, leaving their birth bins depleted relative to the true in situ star formation. The combined effect is that the age distribution in any given radial bin is distorted: peaks are suppressed as stars migrate away, and the wings are broadened due to the influx of migrators.

Star formation also tends to quench earlier in the inner disk, especially in barred systems \citep{khoperskov18,geron24,Ratcliffe2024,Renu2025}. As a result, young stars observed there at late times are more likely to be inward migrators. Conversely, stars that migrate outward early in the disk's history dominate the outer disk at times when in situ star formation was negligible.

A particularly strong exchange occurs across $R\sim1$--$1.4h_d$, which has been shown to be a pivot point in flattening the metallicity gradient in g106 \citep{mcm13}. This is driven by the strong angular momentum exchange across this radius, corresponding to the region just outside the bar's corotation resonance (CR), where stars just inside CR tend to shift outward and those just outside shift inward. This effect is visible in g106: at $lbt<8$~Gyr, the red curves (inner disk) are underestimated by $\sim$30\%, while the yellow--cyan curves (intermediate radii) are overestimated by $\sim$200\%. This is consistent with a roughly symmetric mass exchange across CR, modulated by the exponential surface density of the disk.

The net consequence is that reconstructions of spatially resolved SFHs from present-day stellar radii systematically underestimate early inner-disk star formation (by 25--50\%), with the exception of the very innermost bin ($\sim0.4\,h_d$), which is generally overestimated due to a net inflow from slightly larger radii, though in galaxies with late massive mergers it can instead be underestimated. Conversely, outer-disk star formation is overestimated by more than 200\%. Together, these biases lead to an underestimation of the true rate of inside-out disk growth, with important consequences for measuring disk assembly histories and calibrating galaxy formation models. This is especially relevant for IFU surveys such as CALIFA and MaNGA \citep{gonzalez17}, which often interpret present-day stellar positions as indicators of birth locations.

\subsection{Limitations of age-based migration diagnostics}

Our analysis quantifies the impact of radial migration on SFHs using only stellar ages. This approach is sensitive to net fluxes of stars across radial bins: if stars of a given age leave a bin without being replenished, the local SFR estimate is biased. However, if amount of mass migrate in, the reconstructed SFR–age relation remains unchanged, even though substantial exchange has taken place. In this sense, our estimates represent a lower limit on the full effect of migration. 

Because metallicity gradients exist (e.g., \citealt{anders17b, willett23}), such symmetric exchanges are not neutral chemically: stars of the same age born at smaller radii are more metal-rich than those formed farther out. Thus, to capture the hidden dimension of mixing, chemo-dynamical information must be combined with age-based SFHs. This is one of the goals of the spectroscopic \textit{Gaia} follow-up survey 4MIDABLE-LR \citep{chiappini19}, which targets the Milky Way disk and bulge and will includes detailed chemical information.

\section{Conclusions}

We used a set of zoom-in cosmological simulations to quantify how radial migration affects estimates of the spatially resolved SFH of galactic disks. For each galaxy, we computed the SFH in radial bins using either stellar birth radii or present-day radii, and compared the resulting absolute and fractional differences ($\Delta \mathrm{SFR}$ and $\Delta \mathrm{SFR}_{\mathrm{fr}}$). We find that migration significantly distorts reconstructed SFHs, with systematic biases that depend on both radius and time:

\begin{itemize}
\item Star formation histories inferred from present-day stellar positions systematically underestimate early star formation in the inner disk by 25--50\% and overestimate outer-disk activity at later times by more than 200\%. By contrast, the very central bin ($\sim0.4h_d$) is generally overestimated, owing to a net inflow from slightly larger radii, though in galaxies with late massive mergers it can instead be underestimated. Together, these effects lead to an underestimation of the true rate of inside-out disk growth.

\item Migration suppresses and broadens star formation peaks: subdominant bursts are washed out, while dominant events are redistributed across neighboring radii. Consequently, peaks inferred from stars near the solar neighborhood may have originated elsewhere or at different epochs.

\item A dynamical case study of the Milky Way–like system g106 links these SFH biases directly to the disk response to mergers. Gas-rich and subsequent dry satellites excite lopsidedness, bars, and multi-arm spirals whose Fourier amplitudes peak at the same epochs as the strongest deviations in $\Delta \mathrm{SFR}$ and the nearly symmetric mass exchange between inner and outer radii, consistent with torque-driven radial migration.

\item A particularly strong exchange occurs across $R\sim1$--$1.4h_d$ in the case of barred disks -- just outside bar corotation, where angular momentum transfer drives both SFH biases and metallicity gradient flattening. While barred and non-barred disks show similar average biases in our small sample (cf. \citealt{bernaldez25}), bars can accelerate inner-disk quenching and amplify exchanges near corotation.

\item These results caution against using present-day stellar radii as proxies for birth sites when reconstructing SFHs. They emphasize the need for migration-aware modeling when interpreting IFU surveys such as CALIFA and MaNGA, and for birth-radius estimation methods to recover robust disk assembly histories in the Milky Way.
\end{itemize}

In a forthcoming study \citep{bernaldez25}, we analyze a large sample of Milky Way- and Andromeda-like galaxies from TNG50 \citep{pillepich24} to systematically assess how radial migration biases spatially resolved SFH reconstructions. This work demonstrates that the magnitude and direction of these biases depend strongly on bar strength, disk thickness, and merger history. Further investigation into the connection between galactic structure at the final simulation time and migration history will be valuable for developing corrections to observed SFH measurements in external galaxies.

In the Milky Way, various approaches are possible thanks to the availability of precise chemical, kinematic, and age information from the astrometric mission \textit{Gaia} \citep{gaia18,drimmel23}, combined with existing and upcoming spectroscopic surveys, for example APOGEE \citep{majewski17}, SDSS-V \citep{sdss25}, WEAVE \citep{dalton12}, and 4MOST \citep{dejong19}, and asteroseismic missions, for example K2 \citep{howell14}, TESS \citep{ricker15}, and PLATO \citep{rauer14}. Corrections to SFH estimates at a given radius can be obtained by inferring stellar birth radii from chemical composition and age, as proposed in previous studies (\citealt{minchev18,lu24,ratcliffe23,ratcliffe25}; see also \citealt{frankel18,frankel20,feltzing19,baba25}). This approach is the focus of an upcoming work \citep{ratcliffe26} that applies the \cite{khoperskov25a} orbital superposition method and \cite{ratcliffe25} birth radius estimates to APOGEE data.

\begin{acknowledgements}
We thank the anonymous referee for a thoughtful and constructive report, which led to significant improvements in the clarity and interpretation of our results. IM and BR acknowledge support by the Deutsche Forschungsgemeinschaft under the grant MI 2009/2-1. This work is based on the Master's thesis of K. Attard (2019, Leibniz Institute for Astrophysics Potsdam, AIP).
\end{acknowledgements}

\bibliographystyle{aa}
\bibliography{myreferences}

@ARTICLE{annem24,
       author = {{Annem}, Bhargav and {Khoperskov}, Sergey},
        title = "{Impact of orbiting satellites on star formation rate evolution and metallicity variations in Milky Way-like discs}",
      journal = {\mnras},
         year = 2024,
        month = jan,
       volume = {527},
       number = {2},
        pages = {2426-2436},
          doi = {10.1093/mnras/stad3244},
archivePrefix = {arXiv},
       eprint = {2210.17054},
 primaryClass = {astro-ph.GA},
       adsurl = {https://ui.adsabs.harvard.edu/abs/2024MNRAS.527.2426A}
}

@ARTICLE{desafreitas25,
       author = {{de S{\'a}-Freitas}, Camila and {Gadotti}, Dimitri A. and {Fragkoudi}, Francesca and {Coelho}, Paula and {de Lorenzo-C{\'a}ceres}, Adriana and {Falc{\'o}n-Barroso}, Jes{\'u}s and {S{\'a}nchez-Bl{\'a}zquez}, Patricia and {Kim}, Taehyun and {Mendez-Abreu}, Jairo and {Neumann}, Justus and {Querejeta}, Miguel and {van de Ven}, Glenn},
        title = "{Bar ages derived for the first time in nearby galaxies: Insights into secular evolution from the TIMER sample}",
      journal = {\aap},
         year = 2025,
        month = jun,
       volume = {698},
          eid = {A5},
        pages = {A5},
          doi = {10.1051/0004-6361/202453367},
archivePrefix = {arXiv},
       eprint = {2503.20864},
 primaryClass = {astro-ph.GA},
       adsurl = {https://ui.adsabs.harvard.edu/abs/2025A&A...698A...5D}
}

@ARTICLE{zibetti24,
       author = {{Zibetti}, Stefano and {Rossi}, Edoardo and {Gallazzi}, Anna R.},
        title = "{On the maximum age resolution achievable through stellar population synthesis models}",
      journal = {\mnras},
         year = 2024,
        month = feb,
       volume = {528},
       number = {2},
        pages = {2790-2804},
          doi = {10.1093/mnras/stae178},
archivePrefix = {arXiv},
       eprint = {2401.07335},
 primaryClass = {astro-ph.GA},
       adsurl = {https://ui.adsabs.harvard.edu/abs/2024MNRAS.528.2790Z}
}

@ARTICLE{gonzalez17,
       author = {{Gonz{\'a}lez Delgado}, R.~M. and {P{\'e}rez}, E. and {Cid Fernandes}, R. and {Garc{\'\i}a-Benito}, R. and {L{\'o}pez Fern{\'a}ndez}, R. and {Vale Asari}, N. and {Cortijo-Ferrero}, C. and {de Amorim}, A.~L. and {Lacerda}, E.~A.~D. and {S{\'a}nchez}, S.~F. and {Lehnert}, M.~D. and {Walcher}, C.~J.},
        title = "{Spatially-resolved star formation histories of CALIFA galaxies. Implications for galaxy formation}",
      journal = {\aap},
         year = 2017,
        month = nov,
       volume = {607},
          eid = {A128},
        pages = {A128},
          doi = {10.1051/0004-6361/201730883},
archivePrefix = {arXiv},
       eprint = {1706.06119},
 primaryClass = {astro-ph.GA},
       adsurl = {https://ui.adsabs.harvard.edu/abs/2017A&A...607A.128G}
}

@ARTICLE{peterken20,
       author = {{Peterken}, Thomas and {Merrifield}, Michael and {Arag{\'o}n-Salamanca}, Alfonso and {Fraser-McKelvie}, Amelia and {Avila-Reese}, Vladimir and {Riffel}, Rog{\'e}rio and {Knapen}, Johan and {Drory}, Niv},
        title = "{SDSS-IV MaNGA: Excavating the fossil record of stellar populations in spiral galaxies}",
      journal = {\mnras},
         year = 2020,
        month = jan,
       volume = {495},
       number = {3},
        pages = {3387-3402},
          doi = {10.1093/mnras/staa1303},
archivePrefix = {arXiv},
       eprint = {2005.03012},
 primaryClass = {astro-ph.GA},
       adsurl = {https://ui.adsabs.harvard.edu/abs/2020MNRAS.495.3387P}
}

@ARTICLE{sanchez19,
       author = {{S{\'a}nchez}, S.~F. and {Avila-Reese}, V. and {Rodr{\'\i}guez-Puebla}, A. and {Ibarra-Medel}, H. and {Calette}, R. and {Bershady}, M. and {Hern{\'a}ndez-Toledo}, H. and {Pan}, K. and {Bizyaev}, D.},
        title = "{SDSS-IV MaNGA - an archaeological view of the cosmic star formation history}",
      journal = {\mnras},
         year = 2019,
        month = jan,
       volume = {482},
       number = {2},
        pages = {1557-1586},
          doi = {10.1093/mnras/sty2730},
archivePrefix = {arXiv},
       eprint = {1807.11528},
 primaryClass = {astro-ph.GA},
       adsurl = {https://ui.adsabs.harvard.edu/abs/2019MNRAS.482.1557S}
}

@ARTICLE{Ratcliffe2024,
       author = {{Ratcliffe}, B. and {Khoperskov}, S. and {Minchev}, I. and {Lu}, L. and {de Jong}, R.~S. and {Steinmetz}, M.},
        title = "{Empirical derivation of the metallicity evolution with time and radius using TNG50 Milky Way and Andromeda analogues}",
      journal = {\aap},
         year = 2024,
        month = oct,
       volume = {690},
          eid = {A352},
        pages = {A352},
          doi = {10.1051/0004-6361/202449268},
archivePrefix = {arXiv},
       eprint = {2401.09260},
 primaryClass = {astro-ph.GA},
       adsurl = {https://ui.adsabs.harvard.edu/abs/2024A&A...690A.352R}
}

@ARTICLE{ruiz-lara20,
       author = {{Ruiz-Lara}, Tom{\'a}s and {Gallart}, Carme and {Bernard}, Edouard J. and {Cassisi}, Santi},
        title = "{The recurrent impact of the Sagittarius dwarf on the star formation history of the Milky Way}",
      journal = {Nature Astronomy},
         year = 2020,
        month = may,
       volume = {4},
        pages = {965-973},
          doi = {10.1038/s41550-020-1097-0},
archivePrefix = {arXiv},
       eprint = {2003.12577},
 primaryClass = {astro-ph.GA},
       adsurl = {https://ui.adsabs.harvard.edu/abs/2020NatAs...4..965R}
}

@ARTICLE{Renu2025,
       author = {{Renu}, D. and {Subramanian}, Smitha and {Rao}, Suhasini and {George}, Koshy},
        title = "{Investigating the role of bars in quenching star formation using spatially resolved ultraviolet-optical colour maps}",
      journal = {\aap},
         year = 2025,
        month = apr,
       volume = {696},
          eid = {A118},
        pages = {A118},
          doi = {10.1051/0004-6361/202452701},
archivePrefix = {arXiv},
       eprint = {2501.17075},
 primaryClass = {astro-ph.GA},
       adsurl = {https://ui.adsabs.harvard.edu/abs/2025A&A...696A.118R}
}

@ARTICLE{khoperskov18,
       author = {{Khoperskov}, S. and {Haywood}, M. and {Di Matteo}, P. and {Lehnert}, M.~D. and {Combes}, F.},
        title = "{Bar quenching in gas-rich galaxies}",
      journal = {\aap},
         year = 2018,
        month = jan,
       volume = {609},
          eid = {A60},
        pages = {A60},
          doi = {10.1051/0004-6361/201731211},
archivePrefix = {arXiv},
       eprint = {1709.03604},
 primaryClass = {astro-ph.GA},
       adsurl = {https://ui.adsabs.harvard.edu/abs/2018A&A...609A..60K}
}

@ARTICLE{khoperskov2025_2,
       author = {{Khoperskov}, Sergey and {Steinmetz}, Matthias and {Haywood}, Misha and {van de Ven}, Glenn and {Krajnovi{\'c}}, Davor and {Ratcliffe}, Bridget and {Minchev}, Ivan and {Di Matteo}, Paola and {Kacharov}, Nikolay and {Marques}, L{\'e}a and {Valentini}, Marica and {de Jong}, Roelof S.},
        title = "{Rediscovering the Milky Way with an orbit superposition approach and APOGEE data: II. Chrono-chemo-kinematics of the disc}",
      journal = {\aap},
         year = 2025,
        month = aug,
       volume = {700},
          eid = {A89},
        pages = {A89},
          doi = {10.1051/0004-6361/202453305},
archivePrefix = {arXiv},
       eprint = {2411.16866},
 primaryClass = {astro-ph.GA},
       adsurl = {https://ui.adsabs.harvard.edu/abs/2025A&A...700A..89K}
}

@ARTICLE{khoperskov25a,
       author = {{Khoperskov}, Sergey and {van de Ven}, Glenn and {Steinmetz}, Matthias and {Ratcliffe}, Bridget and {Minchev}, Ivan and {Krajnovi{\'c}}, Davor and {Haywood}, Misha and {Di Matteo}, Paola and {Kacharov}, Nikolay and {Marques}, L{\'e}a and {Valentini}, Marica and {de Jong}, Roelof S.},
        title = "{Rediscovering the Milky Way with an orbit superposition approach and APOGEE data: I. Method validation}",
      journal = {\aap},
         year = 2025,
        month = mar,
       volume = {695},
          eid = {A220},
        pages = {A220},
          doi = {10.1051/0004-6361/202453304},
archivePrefix = {arXiv},
       eprint = {2411.15062},
 primaryClass = {astro-ph.GA},
       adsurl = {https://ui.adsabs.harvard.edu/abs/2025A&A...695A.220K}
}

@ARTICLE{khoperskov23,
       author = {{Khoperskov}, Sergey and {Minchev}, Ivan and {Libeskind}, Noam and {Haywood}, Misha and {Di Matteo}, Paola and {Belokurov}, Vasily and {Steinmetz}, Matthias and {Gomez}, Facundo A. and {Grand}, Robert J.~J. and {Hoffman}, Yehuda and {Knebe}, Alexander and {Sorce}, Jenny G. and {Spaare}, Martin and {Tempel}, Elmo and {Vogelsberger}, Mark},
        title = "{The stellar halo in Local Group Hestia simulations. I. The in situ component and the effect of mergers}",
      journal = {\aap},
         year = 2023,
        month = sep,
       volume = {677},
          eid = {A89},
        pages = {A89},
          doi = {10.1051/0004-6361/202244232},
archivePrefix = {arXiv},
       eprint = {2206.04521},
 primaryClass = {astro-ph.GA},
       adsurl = {https://ui.adsabs.harvard.edu/abs/2023A&A...677A..89K}
}

@ARTICLE{khoperskov20,
       author = {{Khoperskov}, S. and {Di Matteo}, P. and {Haywood}, M. and {G{\'o}mez}, A. and {Snaith}, O.~N.},
        title = "{Escapees from the bar resonances. Presence of low-eccentricity metal-rich stars at the solar vicinity}",
      journal = {\aap},
         year = 2020,
        month = jun,
       volume = {638},
          eid = {A144},
        pages = {A144},
          doi = {10.1051/0004-6361/201937188},
archivePrefix = {arXiv},
       eprint = {1911.12424},
 primaryClass = {astro-ph.GA},
       adsurl = {https://ui.adsabs.harvard.edu/abs/2020A&A...638A.144K}
}

@ARTICLE{anders17b,
   author = {{Anders}, F. and {Chiappini}, C. and {Minchev}, I. and {Miglio}, A. and 
	{Montalb{\'a}n}, J. and {Mosser}, B. and {Rodrigues}, T.~S. and 
	{Santiago}, B.~X. and {Baudin}, F. and {Beers}, T.~C. and {da Costa}, L.~N. and 
	{Garc{\'{\i}}a}, R.~A. and {Garc{\'{\i}}a-Hern{\'a}ndez}, D.~A. and 
	{Holtzman}, J. and {Maia}, M.~A.~G. and {Majewski}, S. and {Mathur}, S. and 
	{Noels-Grotsch}, A. and {Pan}, K. and {Schneider}, D.~P. and 
	{Schultheis}, M. and {Steinmetz}, M. and {Valentini}, M. and 
	{Zamora}, O.},
    title = "{Red giants observed by CoRoT and APOGEE: The evolution of the Milky Way's radial metallicity gradient}",
  journal = {\aap},
archivePrefix = "arXiv",
   eprint = {1608.04951},
     year = 2017,
    month = apr,
   volume = 600,
      eid = {A70},
    pages = {A70},
      doi = {10.1051/0004-6361/201629363},
   adsurl = {http://adsabs.harvard.edu/abs/2017A%26A...600A..70A}
}

@ARTICLE{aparicio09,
       author = {{Aparicio}, Antonio and {Hidalgo}, Sebastian L.},
        title = "{IAC-pop: Finding the Star Formation History of Resolved Galaxies}",
      journal = {\aj},
         year = "2009",
        month = "Aug",
       volume = {138},
       number = {2},
        pages = {558-567},
          doi = {10.1088/0004-6256/138/2/558},
archivePrefix = {arXiv},
       eprint = {0906.0712},
 primaryClass = {astro-ph.CO},
       adsurl = {https://ui.adsabs.harvard.edu/abs/2009AJ....138..558A}
}

@ARTICLE{baba25,
       author = {{Baba}, Junichi},
        title = "{Influence of bar formation on star formation segregation and stellar migration: Implications for variations in the age distribution of Milky Way disk stars}",
      journal = {\pasj},
         year = 2025,
        month = aug,
       volume = {77},
       number = {4},
        pages = {916-923},
          doi = {10.1093/pasj/psaf062},
archivePrefix = {arXiv},
       eprint = {2505.16528},
 primaryClass = {astro-ph.GA},
       adsurl = {https://ui.adsabs.harvard.edu/abs/2025PASJ...77..916B}
}

@ARTICLE{bernaldez25,
       author = {{Bernaldez}, J.~P. and {Minchev}, I. and {Ratcliffe}, B. and {Marques}, L. and {Sysoliatina}, K. and {Walcher}, J. and {Khoperskov}, S. and {Martig}, M. and {de Jong}, R. and {Steinmetz}, M.},
        title = "{The Impact of Radial Migration on Disk Galaxy SFHs: II. The Role of bar strength, disk thickness, and merger history}",
      journal = {A\&A, submitted},
         year = 2025,
        month = aug,
          eid = {arXiv:2508.19340},
        pages = {arXiv:2508.19340},
archivePrefix = {arXiv},
       eprint = {2508.19340},
 primaryClass = {astro-ph.GA},
       adsurl = {https://ui.adsabs.harvard.edu/abs/2025arXiv250819340B}
}

@ARTICLE{BC02,
       author = {{Bournaud}, F. and {Combes}, F.},
        title = "{Gas accretion on spiral galaxies: Bar formation and renewal}",
      journal = {\aap},
         year = 2002,
        month = sep,
       volume = {392},
        pages = {83-102},
          doi = {10.1051/0004-6361:20020920},
archivePrefix = {arXiv},
       eprint = {astro-ph/0206273},
 primaryClass = {astro-ph},
       adsurl = {https://ui.adsabs.harvard.edu/abs/2002A&A...392...83B}
}

@ARTICLE{bernard15,
       author = {{Bernard}, Edouard J. and {Ferguson}, Annette M.~N. and {Chapman}, Scott C. and {Ibata}, Rodrigo A. and {Irwin}, Mike J. and {Lewis}, Geraint F. and {McConnachie}, Alan W.},
        title = "{The spatially-resolved star formation history of the M31 outer disc}",
      journal = {\mnras},
         year = 2015,
        month = oct,
       volume = {453},
       number = {1},
        pages = {L113-L117},
          doi = {10.1093/mnrasl/slv116},
archivePrefix = {arXiv},
       eprint = {1508.01559},
 primaryClass = {astro-ph.GA},
       adsurl = {https://ui.adsabs.harvard.edu/abs/2015MNRAS.453L.113B}
}

@ARTICLE{bernard18,
       author = {{Bernard}, Edouard J. and {Schultheis}, Mathias and {Di Matteo}, Paola and
         {Hill}, Vanessa and {Haywood}, Misha and {Calamida}, Annalisa},
        title = "{Star formation history of the Galactic bulge from deep HST imaging of low reddening windows}",
      journal = {\mnras},
         year = "2018",
        month = "Jul",
       volume = {477},
       number = {3},
        pages = {3507-3519},
          doi = {10.1093/mnras/sty902},
archivePrefix = {arXiv},
       eprint = {1801.01426},
 primaryClass = {astro-ph.GA},
       adsurl = {https://ui.adsabs.harvard.edu/abs/2018MNRAS.477.3507B}
}

@ARTICLE{brunetti11,
   author = {{Brunetti}, M. and {Chiappini}, C. and {Pfenniger}, D.},
    title = "{Stellar diffusion in barred spiral galaxies}",
  journal = {\aap},
archivePrefix = "arXiv",
   eprint = {1108.5631},
 primaryClass = "astro-ph.GA",
     year = 2011,
    month = oct,
   volume = 534,
      eid = {A75},
    pages = {A75},
      doi = {10.1051/0004-6361/201117566},
   adsurl = {http://cdsads.u-strasbg.fr/abs/2011A%26A...534A..75B}
}

@ARTICLE{cohen24,
       author = {{Cohen}, Roger E. and {McQuinn}, Kristen B.~W. and {Murray}, Claire E. and {Williams}, Benjamin F. and {Choi}, Yumi and {Lindberg}, Christina W. and {Burhenne}, Clare and {Gordon}, Karl D. and {Yanchulova Merica-Jones}, Petia and {Gilbert}, Karoline M. and {Boyer}, Martha L. and {Goldman}, Steven and {Dolphin}, Andrew E. and {Telford}, O. Grace},
        title = "{Scylla. II. The Spatially Resolved Star Formation History of the Large Magellanic Cloud Reveals an Inverted Radial Age Gradient}",
      journal = {\apj},
         year = 2024,
        month = nov,
       volume = {975},
       number = {1},
          eid = {42},
        pages = {42},
          doi = {10.3847/1538-4357/ad6cd5},
archivePrefix = {arXiv},
       eprint = {2410.11696},
 primaryClass = {astro-ph.GA},
       adsurl = {https://ui.adsabs.harvard.edu/abs/2024ApJ...975...42C}
}

@ARTICLE{carrillo18,
       author = {{Carrillo}, I. and {Minchev}, I. and {Kordopatis}, G. and
         {Steinmetz}, M. and {Binney}, J. and {Anders}, F. and
         {Bienaym{\'e}}, O. and {Bland-Hawthorn}, J. and {Famaey}, B. and
         {Freeman}, K.~C. and {Gilmore}, G. and {Gibson}, B.~K. and
         {Grebel}, E.~K. and {Helmi}, A. and {Just}, A. and {Kunder}, A. and
         {McMillan}, P. and {Monari}, G. and {Munari}, U. and {Navarro}, J. and
         {Parker}, Q.~A. and {Reid}, W. and {Seabroke}, G. and {Sharma}, S. and
         {Siebert}, A. and {Watson}, F. and {Wojno}, J. and {Wyse}, R.~F.~G. and
         {Zwitter}, T.},
        title = "{Is the Milky Way still breathing? RAVE-Gaia streaming motions}",
      journal = {\mnras},
         year = "2018",
        month = "Apr",
       volume = {475},
       number = {2},
        pages = {2679-2696},
          doi = {10.1093/mnras/stx3342},
archivePrefix = {arXiv},
       eprint = {1710.03763},
 primaryClass = {astro-ph.GA},
       adsurl = {https://ui.adsabs.harvard.edu/abs/2018MNRAS.475.2679C}
}

@ARTICLE{cid13,
author = {{Cid Fernandes}, R. and {P{\'e}rez}, E. and {Garc{\'\i}a Benito}, R. and
{Gonz{\'a}lez Delgado}, R.~M. and {de Amorim}, A.~L. and {S{\'a}nchez}, S.~F. and {Husemann}, B. and {Falc{\'o}n Barroso}, J. and {S{\'a}nchez-Bl{\'a}zquez}, P. and {Walcher}, C.~J. and {Mast}, D.},
title = "{Resolving galaxies in time and space. I. Applying STARLIGHT to CALIFA datacubes}", journal = {\aap},
year = "2013", month = "Sep", volume = {557},
eid = {A86}, pages = {A86},
doi = {10.1051/0004-6361/201220616}, archivePrefix = {arXiv},
eprint = {1304.5788}, primaryClass = {astro-ph.CO},
adsurl = {https://ui.adsabs.harvard.edu/abs/2013A&A...557A..86C}
}

@ARTICLE{chiappini19,
       author = {{Chiappini}, C. and {Minchev}, I. and {Starkenburg}, E. and {Anders}, F. and {Gentile Fusillo}, N. and {Gerhard}, O. and {Guiglion}, G. and {Khalatyan}, A. and {Kordopatis}, G. and {Lemasle}, B. and {Matijevic}, G. and {Queiroz}, A.~B.~D.~A. and {Schwope}, A. and {Steinmetz}, M. and {Storm}, J. and {Traven}, G. and {Tremblay}, P. -E. and {Valentini}, M. and {Andrae}, R. and {Arentsen}, A. and {Asplund}, M. and {Bensby}, T. and {Bergemann}, M. and {Casagrande}, L. and {Church}, R. and {Cescutti}, G. and {Feltzing}, S. and {Fouesneau}, M. and {Grebel}, E.~K. and {Kovalev}, M. and {McMillan}, P. and {Monari}, G. and {Rybizki}, J. and {Ryde}, N. and {Rix}, H. -W. and {Walton}, N. and {Xiang}, M. and {Zucker}, D. and {4MIDABLE-Lr Team}},
        title = "{4MOST Consortium Survey 3: Milky Way Disc and Bulge Low-Resolution Survey (4MIDABLE-LR)}",
      journal = {The Messenger},
         year = 2019,
        month = mar,
       volume = {175},
        pages = {30-34},
          doi = {10.18727/0722-6691/5122},
archivePrefix = {arXiv},
       eprint = {1903.02469},
 primaryClass = {astro-ph.GA},
       adsurl = {https://ui.adsabs.harvard.edu/abs/2019Msngr.175...30C}
}

@ARTICLE{chiappini97,
       author = {{Chiappini}, C. and {Matteucci}, F. and {Gratton}, R.},
        title = "{The Chemical Evolution of the Galaxy: The Two-Infall Model}",
      journal = {\apj},
         year = 1997,
        month = mar,
       volume = {477},
       number = {2},
        pages = {765-780},
          doi = {10.1086/303726},
archivePrefix = {arXiv},
       eprint = {astro-ph/9609199},
 primaryClass = {astro-ph},
       adsurl = {https://ui.adsabs.harvard.edu/abs/1997ApJ...477..765C}
}

@ARTICLE{dantas25,
       author = {{Dantas}, M.~L.~L. and {Smiljanic}, R. and {de Souza}, R.~S. and {Tissera}, P.~B. and {Magrini}, L.},
        title = "{Probing the origins: I. Generalised additive model inference of birth radii for Milky Way stars in the solar vicinity}",
      journal = {\aap},
         year = 2025,
        month = apr,
       volume = {696},
          eid = {A205},
        pages = {A205},
          doi = {10.1051/0004-6361/202453034},
archivePrefix = {arXiv},
       eprint = {2502.20441},
 primaryClass = {astro-ph.GA},
       adsurl = {https://ui.adsabs.harvard.edu/abs/2025A&A...696A.205D}
}

@ARTICLE{delalc25,
       author = {{del Alc{\'a}zar-Juli{\`a}}, M. and {Figueras}, F. and {Robin}, A.~C. and {Bienaym{\'e}}, O. and {Anders}, F.},
        title = "{Joint inference of the Milky Way's star formation history and initial mass function from Gaia all-sky G < 13 data}",
      journal = {\aap},
         year = 2025,
        month = may,
       volume = {697},
          eid = {A128},
        pages = {A128},
          doi = {10.1051/0004-6361/202453606},
archivePrefix = {arXiv},
       eprint = {2501.17236},
 primaryClass = {astro-ph.GA},
       adsurl = {https://ui.adsabs.harvard.edu/abs/2025A&A...697A.128D}
}

@ARTICLE{dicintio21,
       author = {{Di Cintio}, Arianna and {Mostoghiu}, Robert and {Knebe}, Alexander and {Navarro}, Julio F.},
        title = "{Pericentric passage-driven star formation in satellite galaxies and their hosts: CLUES from local group simulations}",
      journal = {\mnras},
         year = 2021,
        month = sep,
       volume = {506},
       number = {1},
        pages = {531-545},
          doi = {10.1093/mnras/stab1682},
archivePrefix = {arXiv},
       eprint = {2103.02739},
 primaryClass = {astro-ph.GA},
       adsurl = {https://ui.adsabs.harvard.edu/abs/2021MNRAS.506..531D}
}

@INPROCEEDINGS{dalton12,
   author = {{Dalton}, G. and {Trager}, S.~C. and {Abrams}, D.~C. and {Carter}, D. and 
	{Bonifacio}, P. and {Aguerri}, J.~A.~L. and {MacIntosh}, M. and 
	{Evans}, C. and {Lewis}, I. and {Navarro}, R. and {Agocs}, T. and 
	{Dee}, K. and {Rousset}, S. and {Tosh}, I. and {Middleton}, K. and 
	{Pragt}, J. and {Terrett}, D. and {Brock}, M. and {Benn}, C. and 
	{Verheijen}, M. and {Cano Infantes}, D. and {Bevil}, C. and 
	{Steele}, I. and {Mottram}, C. and {Bates}, S. and {Gribbin}, F.~J. and 
	{Rey}, J. and {Rodriguez}, L.~F. and {Delgado}, J.~M. and {Guinouard}, I. and 
	{Walton}, N. and {Irwin}, M.~J. and {Jagourel}, P. and {Stuik}, R. and 
	{Gerlofsma}, G. and {Roelfsma}, R. and {Skillen}, I. and {Ridings}, A. and 
	{Balcells}, M. and {Daban}, J.-B. and {Gouvret}, C. and {Venema}, L. and 
	{Girard}, P.},
    title = "{WEAVE: the next generation wide-field spectroscopy facility for the William Herschel Telescope}",
booktitle = {Ground-based and Airborne Instrumentation for Astronomy IV},
     year = 2012,
   series = {\procspie},
   volume = 8446,
    month = sep,
      eid = {84460P},
    pages = {84460P},
      doi = {10.1117/12.925950},
   adsurl = {http://adsabs.harvard.edu/abs/2012SPIE.8446E..0PD}
}

@ARTICLE{debattista06,
   author = {{Debattista}, V.~P. and {Mayer}, L. and {Carollo}, C.~M. and 
	{Moore}, B. and {Wadsley}, J. and {Quinn}, T.},
    title = "{The Secular Evolution of Disk Structural Parameters}",
  journal = {\apj},
   eprint = {arXiv:astro-ph/0509310},
     year = 2006,
    month = jul,
   volume = 645,
    pages = {209-227},
      doi = {10.1086/504147},
   adsurl = {http://adsabs.harvard.edu/abs/2006ApJ...645..209D}
}

@ARTICLE{khoperskov19,
       author = {{Khoperskov}, Sergey and {Di Matteo}, Paola and {Gerhard}, Ortwin and {Katz}, David and {Haywood}, Misha and {Combes}, Fran{\c{c}}oise and {Berczik}, Peter and {Gomez}, Ana},
        title = "{The echo of the bar buckling: Phase-space spirals in Gaia Data Release 2}",
      journal = {\aap},
         year = 2019,
        month = feb,
       volume = {622},
          eid = {L6},
        pages = {L6},
          doi = {10.1051/0004-6361/201834707},
       adsurl = {https://ui.adsabs.harvard.edu/abs/2019A&A...622L...6K}
}

@ARTICLE{snaith15,
       author = {{Snaith}, O. and {Haywood}, M. and {Di Matteo}, P. and {Lehnert}, M.~D. and {Combes}, F. and {Katz}, D. and {G{\'o}mez}, A.},
        title = "{Reconstructing the star formation history of the Milky Way disc(s) from chemical abundances}",
      journal = {\aap},
         year = 2015,
        month = jun,
       volume = {578},
          eid = {A87},
        pages = {A87},
          doi = {10.1051/0004-6361/201424281},
archivePrefix = {arXiv},
       eprint = {1410.3829},
 primaryClass = {astro-ph.GA},
       adsurl = {https://ui.adsabs.harvard.edu/abs/2015A&A...578A..87S}
}

@ARTICLE{dejong19,
       author = {{de Jong}, R.~S. and {Agertz}, O. and {Berbel}, A.~A. and {Aird}, J. and {Alexander}, D.~A. and {Amarsi}, A. and {Anders}, F. and {Andrae}, R. and {Ansarinejad}, B. and {Ansorge}, W. and {Antilogus}, P. and {Anwand-Heerwart}, H. and {Arentsen}, A. and {Arnadottir}, A. and {Asplund}, M. and {Auger}, M. and {Azais}, N. and {Baade}, D. and {Baker}, G. and {Baker}, S. and {Balbinot}, E. and {Baldry}, I.~K. and {Banerji}, M. and {Barden}, S. and {Barklem}, P. and {Barth{\'e}l{\'e}my-Mazot}, E. and {Battistini}, C. and {Bauer}, S. and {Bell}, C.~P.~M. and {Bellido-Tirado}, O. and {Bellstedt}, S. and {Belokurov}, V. and {Bensby}, T. and {Bergemann}, M. and {Bestenlehner}, J.~M. and {Bielby}, R. and {Bilicki}, M. and {Blake}, C. and {Bland-Hawthorn}, J. and {Boeche}, C. and {Boland}, W. and {Boller}, T. and {Bongard}, S. and {Bongiorno}, A. and {Bonifacio}, P. and {Boudon}, D. and {Brooks}, D. and {Brown}, M.~J.~I. and {Brown}, R. and {Br{\"u}ggen}, M. and {Brynnel}, J. and {Brzeski}, J. and {Buchert}, T. and {Buschkamp}, P. and {Caffau}, E. and {Caillier}, P. and {Carrick}, J. and {Casagrande}, L. and {Case}, S. and {Casey}, A. and {Cesarini}, I. and {Cescutti}, G. and {Chapuis}, D. and {Chiappini}, C. and {Childress}, M. and {Christlieb}, N. and {Church}, R. and {Cioni}, M.-R.~L. and {Cluver}, M. and {Colless}, M. and {Collett}, T. and {Comparat}, J. and {Cooper}, A. and {Couch}, W. and {Courbin}, F. and {Croom}, S. and {Croton}, D. and {Daguis{\'e}}, E. and {Dalton}, G. and {Davies}, L.~J.~M. and {Davis}, T. and {de Laverny}, P. and {Deason}, A. and {Dionies}, F. and {Disseau}, K. and {Doel}, P. and {D{\"o}scher}, D. and {Driver}, S.~P. and {Dwelly}, T. and {Eckert}, D. and {Edge}, A. and {Edvardsson}, B. and {Youssoufi}, D.~E. and {Elhaddad}, A. and {Enke}, H. and {Erfanianfar}, G. and {Farrell}, T. and {Fechner}, T. and {Feiz}, C. and {Feltzing}, S. and {Ferreras}, I. and {Feuerstein}, D. and {Feuillet}, D. and {Finoguenov}, A. and {Ford}, D. and {Fotopoulou}, S. and {Fouesneau}, M. and {Frenk}, C. and {Frey}, S. and {Gaessler}, W. and {Geier}, S. and {Gentile Fusillo}, N. and {Gerhard}, O. and {Giannantonio}, T. and {Giannone}, D. and {Gibson}, B. and {Gillingham}, P. and {Gonz{\'a}lez-Fern{\'a}ndez}, C. and {Gonzalez-Solares}, E. and {Gottloeber}, S. and {Gould}, A. and {Grebel}, E.~K. and {Gueguen}, A. and {Guiglion}, G. and {Haehnelt}, M. and {Hahn}, T. and {Hansen}, C.~J. and {Hartman}, H. and {Hauptner}, K. and {Hawkins}, K. and {Haynes}, D. and {Haynes}, R. and {Heiter}, U. and {Helmi}, A. and {Aguayo}, C.~H. and {Hewett}, P. and {Hinton}, S. and {Hobbs}, D. and {Hoenig}, S. and {Hofman}, D. and {Hook}, I. and {Hopgood}, J. and {Hopkins}, A. and {Hourihane}, A. and {Howes}, L. and {Howlett}, C. and {Huet}, T. and {Irwin}, M. and {Iwert}, O. and {Jablonka}, P. and {Jahn}, T. and {Jahnke}, K. and {Jarno}, A. and {Jin}, S. and {Jofre}, P. and {Johl}, D. and {Jones}, D. and {J{\"o}nsson}, H. and {Jordan}, C. and {Karovicova}, I. and {Khalatyan}, A. and {Kelz}, A. and {Kennicutt}, R. and {King}, D. and {Kitaura}, F. and {Klar}, J. and {Klauser}, U. and {Kneib}, J.-P. and {Koch}, A. and {Koposov}, S. and {Kordopatis}, G. and {Korn}, A. and {Kosmalski}, J. and {Kotak}, R. and {Kovalev}, M. and {Kreckel}, K. and {Kripak}, Y. and {Krumpe}, M. and {Kuijken}, K. and {Kunder}, A. and {Kushniruk}, I. and {Lam}, M.~I. and {Lamer}, G. and {Laurent}, F. and {Lawrence}, J. and {Lehmitz}, M. and {Lemasle}, B. and {Lewis}, J. and {Li}, B. and {Lidman}, C. and {Lind}, K. and {Liske}, J. and {Lizon}, J.-L. and {Loveday}, J. and {Ludwig}, H.-G. and {McDermid}, R.~M. and {Maguire}, K. and {Mainieri}, V. and {Mali}, S. and {Mandel}, H.},
        title = "{4MOST: Project overview and information for the First Call for Proposals}",
      journal = {The Messenger},
         year = 2019,
        month = mar,
       volume = {175},
        pages = {3-11},
          doi = {10.18727/0722-6691/5117},
archivePrefix = {arXiv},
       eprint = {1903.02464},
 primaryClass = {astro-ph.IM},
       adsurl = {https://ui.adsabs.harvard.edu/abs/2019Msngr.175....3D}
}

@ARTICLE{drimmel23,
       author = {{Gaia Collaboration} and {Drimmel}, R. and {Romero-G{\'o}mez}, M. and {Chemin}, L. and {Ramos}, P. and {Poggio}, E. and {Ripepi}, V. and {Andrae}, R. and {Blomme}, R. and {Cantat-Gaudin}, T. and {Castro-Ginard}, A. and {Clementini}, G. and {Figueras}, F. and {Fouesneau}, M. and {Fr{\'e}mat}, Y. and {Jardine}, K. and {Khanna}, S. and {Lobel}, A. and {Marshall}, D.~J. and {Muraveva}, T. and {Brown}, A.~G.~A. and {Vallenari}, A. and {Prusti}, T. and {de Bruijne}, J.~H.~J. and {Arenou}, F. and {Babusiaux}, C. and {Biermann}, M. and {Creevey}, O.~L. and {Ducourant}, C. and {Evans}, D.~W. and {Eyer}, L. and {Guerra}, R. and {Hutton}, A. and {Jordi}, C. and {Klioner}, S.~A. and {Lammers}, U.~L. and {Lindegren}, L. and {Luri}, X. and {Mignard}, F. and {Panem}, C. and {Pourbaix}, D. and {Randich}, S. and {Sartoretti}, P. and {Soubiran}, C. and {Tanga}, P. and {Walton}, N.~A. and {Bailer-Jones}, C.~A.~L. and {Bastian}, U. and {Jansen}, F. and {Katz}, D. and {Lattanzi}, M.~G. and {van Leeuwen}, F. and {Bakker}, J. and {Cacciari}, C. and {Casta{\~n}eda}, J. and {De Angeli}, F. and {Fabricius}, C. and {Galluccio}, L. and {Guerrier}, A. and {Heiter}, U. and {Masana}, E. and {Messineo}, R. and {Mowlavi}, N. and {Nicolas}, C. and {Nienartowicz}, K. and {Pailler}, F. and {Panuzzo}, P. and {Riclet}, F. and {Roux}, W. and {Seabroke}, G.~M. and {Sordo}, R. and {Th{\'e}venin}, F. and {Gracia-Abril}, G. and {Portell}, J. and {Teyssier}, D. and {Altmann}, M. and {Audard}, M. and {Bellas-Velidis}, I. and {Benson}, K. and {Berthier}, J. and {Burgess}, P.~W. and {Busonero}, D. and {Busso}, G. and {C{\'a}novas}, H. and {Carry}, B. and {Cellino}, A. and {Cheek}, N. and {Damerdji}, Y. and {Davidson}, M. and {de Teodoro}, P. and {Nu{\~n}ez Campos}, M. and {Delchambre}, L. and {Dell'Oro}, A. and {Esquej}, P. and {Fern{\'a}ndez-Hern{\'a}ndez}, J. and {Fraile}, E. and {Garabato}, D. and {Garc{\'\i}a-Lario}, P. and {Gosset}, E. and {Haigron}, R. and {Halbwachs}, J. -L. and {Hambly}, N.~C. and {Harrison}, D.~L. and {Hern{\'a}ndez}, J. and {Hestroffer}, D. and {Hodgkin}, S.~T. and {Holl}, B. and {Jan{\ss}en}, K. and {Jevardat de Fombelle}, G. and {Jordan}, S. and {Krone-Martins}, A. and {Lanzafame}, A.~C. and {L{\"o}ffler}, W. and {Marchal}, O. and {Marrese}, P.~M. and {Moitinho}, A. and {Muinonen}, K. and {Osborne}, P. and {Pancino}, E. and {Pauwels}, T. and {Recio-Blanco}, A. and {Reyl{\'e}}, C. and {Riello}, M. and {Rimoldini}, L. and {Roegiers}, T. and {Rybizki}, J. and {Sarro}, L.~M. and {Siopis}, C. and {Smith}, M. and {Sozzetti}, A. and {Utrilla}, E. and {van Leeuwen}, M. and {Abbas}, U. and {{\'A}brah{\'a}m}, P. and {Abreu Aramburu}, A. and {Aerts}, C. and {Aguado}, J.~J. and {Ajaj}, M. and {Aldea-Montero}, F. and {Altavilla}, G. and {{\'A}lvarez}, M.~A. and {Alves}, J. and {Anders}, F. and {Anderson}, R.~I. and {Anglada Varela}, E. and {Antoja}, T. and {Baines}, D. and {Baker}, S.~G. and {Balaguer-N{\'u}{\~n}ez}, L. and {Balbinot}, E. and {Balog}, Z. and {Barache}, C. and {Barbato}, D. and {Barros}, M. and {Barstow}, M.~A. and {Bartolom{\'e}}, S. and {Bassilana}, J. -L. and {Bauchet}, N. and {Becciani}, U. and {Bellazzini}, M. and {Berihuete}, A. and {Bernet}, M. and {Bertone}, S. and {Bianchi}, L. and {Binnenfeld}, A. and {Blanco-Cuaresma}, S. and {Boch}, T. and {Bombrun}, A. and {Bossini}, D. and {Bouquillon}, S. and {Bragaglia}, A. and {Bramante}, L. and {Breedt}, E. and {Bressan}, A. and {Brouillet}, N. and {Brugaletta}, E. and {Bucciarelli}, B. and {Burlacu}, A. and {Butkevich}, A.~G. and {Buzzi}, R. and {Caffau}, E. and {Cancelliere}, R. and {Carballo}, R. and {Carlucci}, T. and {Carnerero}, M.~I. and {Carrasco}, J.~M. and {Casamiquela}, L. and {Castellani}, M. and {Chaoul}, L. and {Charlot}, P. and {Chiaramida}, V. and {Chiavassa}, A. and {Chornay}, N. and {Comoretto}, G. and {Contursi}, G. and {Cooper}, W.~J. and {Cornez}, T. and {Cowell}, S. and {Crifo}, F. and {Cropper}, M. and {Crosta}, M. and {Crowley}, C. and {Dafonte}, C. and {Dapergolas}, A. and {David}, P. and {de Laverny}, P. and {De Luise}, F. and {De March}, R. and {De Ridder}, J. and {de Souza}, R. and {de Torres}, A. and {del Peloso}, E.~F. and {del Pozo}, E. and {Delbo}, M. and {Delgado}, A. and {Delisle}, J. -B. and {Demouchy}, C. and {Dharmawardena}, T.~E. and {Di Matteo}, P. and {Diakite}, S. and {Diener}, C. and {Distefano}, E. and {Dolding}, C. and {Enke}, H. and {Fabre}, C. and {Fabrizio}, M. and {Faigler}, S. and {Fedorets}, G. and {Fernique}, P. and {Fournier}, Y. and {Fouron}, C. and {Fragkoudi}, F. and {Gai}, M. and {Garcia-Gutierrez}, A. and {Garcia-Reinaldos}, M. and {Garc{\'\i}a-Torres}, M. and {Garofalo}, A. and {Gavel}, A. and {Gavras}, P. and {Gerlach}, E. and {Geyer}, R. and {Giacobbe}, P. and {Gilmore}, G. and {Girona}, S. and {Giuffrida}, G. and {Gomel}, R. and {Gomez}, A. and {Gonz{\'a}lez-N{\'u}{\~n}ez}, J. and {Gonz{\'a}lez-Santamar{\'\i}a}, I. and {Gonz{\'a}lez-Vidal}, J.~J. and {Granvik}, M. and {Guillout}, P. and {Guiraud}, J. and {Guti{\'e}rrez-S{\'a}nchez}, R. and {Guy}, L.~P. and {Hatzidimitriou}, D. and {Hauser}, M. and {Haywood}, M. and {Helmer}, A. and {Helmi}, A. and {Sarmiento}, M.~H. and {Hidalgo}, S.~L. and {H{\l}adczuk}, N. and {Hobbs}, D. and {Holland}, G. and {Huckle}, H.~E. and {Jasniewicz}, G. and {Jean-Antoine Piccolo}, A. and {Jim{\'e}nez-Arranz}, {\'O}. and {Juaristi Campillo}, J. and {Julbe}, F. and {Karbevska}, L. and {Kervella}, P. and {Kordopatis}, G. and {Korn}, A.~J. and {K{\'o}sp{\'a}l}, {\'A}. and {Kostrzewa-Rutkowska}, Z. and {Kruszy{\'n}ska}, K. and {Kun}, M. and {Laizeau}, P. and {Lambert}, S. and {Lanza}, A.~F. and {Lasne}, Y. and {Le Campion}, J. -F. and {Lebreton}, Y. and {Lebzelter}, T. and {Leccia}, S. and {Leclerc}, N. and {Lecoeur-Taibi}, I. and {Liao}, S. and {Licata}, E.~L. and {Lindstr{\o}m}, H.~E.~P. and {Lister}, T.~A. and {Livanou}, E. and {Lorca}, A. and {Loup}, C. and {Madrero Pardo}, P. and {Magdaleno Romeo}, A. and {Managau}, S. and {Mann}, R.~G. and {Manteiga}, M. and {Marchant}, J.~M. and {Marconi}, M. and {Marcos}, J. and {Marcos Santos}, M.~M.~S. and {Mar{\'\i}n Pina}, D. and {Marinoni}, S. and {Marocco}, F. and {Martin Polo}, L. and {Mart{\'\i}n-Fleitas}, J.~M. and {Marton}, G. and {Mary}, N. and {Masip}, A. and {Massari}, D. and {Mastrobuono-Battisti}, A. and {Mazeh}, T. and {McMillan}, P.~J. and {Messina}, S. and {Michalik}, D. and {Millar}, N.~R. and {Mints}, A. and {Molina}, D. and {Molinaro}, R. and {Moln{\'a}r}, L. and {Monari}, G. and {Mongui{\'o}}, M. and {Montegriffo}, P. and {Montero}, A. and {Mor}, R. and {Mora}, A. and {Morbidelli}, R. and {Morel}, T. and {Morris}, D. and {Murphy}, C.~P. and {Musella}, I. and {Nagy}, Z. and {Noval}, L. and {Oca{\~n}a}, F. and {Ogden}, A. and {Ordenovic}, C. and {Osinde}, J.~O. and {Pagani}, C. and {Pagano}, I. and {Palaversa}, L. and {Palicio}, P.~A. and {Pallas-Quintela}, L. and {Panahi}, A. and {Payne-Wardenaar}, S. and {Pe{\~n}alosa Esteller}, X. and {Penttil{\"a}}, A. and {Pichon}, B. and {Piersimoni}, A.~M. and {Pineau}, F. -X. and {Plachy}, E. and {Plum}, G. and {Pr{\v{s}}a}, A. and {Pulone}, L. and {Racero}, E. and {Ragaini}, S. and {Rainer}, M. and {Raiteri}, C.~M. and {Ramos-Lerate}, M. and {Re Fiorentin}, P. and {Regibo}, S. and {Richards}, P.~J. and {Rios Diaz}, C. and {Riva}, A. and {Rix}, H. -W. and {Rixon}, G. and {Robichon}, N. and {Robin}, A.~C. and {Robin}, C. and {Roelens}, M. and {Rogues}, H.~R.~O. and {Rohrbasser}, L. and {Rowell}, N. and {Royer}, F. and {Ruz Mieres}, D. and {Rybicki}, K.~A. and {Sadowski}, G. and {S{\'a}ez N{\'u}{\~n}ez}, A. and {Sagrist{\`a} Sell{\'e}s}, A. and {Sahlmann}, J. and {Salguero}, E. and {Samaras}, N. and {Sanchez Gimenez}, V. and {Sanna}, N. and {Santove{\~n}a}, R. and {Sarasso}, M. and {Schultheis}, M.~S. and {Sciacca}, E. and {Segol}, M. and {Segovia}, J.~C. and {S{\'e}gransan}, D. and {Semeux}, D. and {Shahaf}, S. and {Siddiqui}, H.~I. and {Siebert}, A. and {Siltala}, L. and {Silvelo}, A. and {Slezak}, E. and {Slezak}, I. and {Smart}, R.~L. and {Snaith}, O.~N. and {Solano}, E. and {Solitro}, F. and {Souami}, D. and {Souchay}, J. and {Spagna}, A. and {Spina}, L. and {Spoto}, F. and {Steele}, I.~A. and {Steidelm{\"u}ller}, H. and {Stephenson}, C.~A. and {S{\"u}veges}, M. and {Surdej}, J. and {Szabados}, L. and {Szegedi-Elek}, E. and {Taris}, F. and {Taylor}, M.~B. and {Teixeira}, R. and {Tolomei}, L. and {Tonello}, N. and {Torra}, F. and {Torra}, J. and {Torralba Elipe}, G. and {Trabucchi}, M. and {Tsounis}, A.~T. and {Turon}, C. and {Ulla}, A. and {Unger}, N. and {Vaillant}, M.~V. and {van Dillen}, E. and {van Reeven}, W. and {Vanel}, O. and {Vecchiato}, A. and {Viala}, Y. and {Vicente}, D. and {Voutsinas}, S. and {Weiler}, M. and {Wevers}, T. and {Wyrzykowski}, {\L}. and {Yoldas}, A. and {Yvard}, P. and {Zhao}, H. and {Zorec}, J. and {Zucker}, S. and {Zwitter}, T.},
        title = "{Gaia Data Release 3. Mapping the asymmetric disc of the Milky Way}",
      journal = {\aap},
         year = 2023,
        month = jun,
       volume = {674},
          eid = {A37},
        pages = {A37},
          doi = {10.1051/0004-6361/202243797},
archivePrefix = {arXiv},
       eprint = {2206.06207},
 primaryClass = {astro-ph.GA},
       adsurl = {https://ui.adsabs.harvard.edu/abs/2023A&A...674A..37G}
}

@ARTICLE{elmegreen16,
       author = {{Elmegreen}, Bruce G. and {Struck}, Curtis},
        title = "{Exponential Disks from Stellar Scattering. III. Stochastic Models}",
      journal = {\apj},
         year = 2016,
        month = oct,
       volume = {830},
       number = {2},
          eid = {115},
        pages = {115},
          doi = {10.3847/0004-637X/830/2/115},
archivePrefix = {arXiv},
       eprint = {1607.07595},
 primaryClass = {astro-ph.GA},
       adsurl = {https://ui.adsabs.harvard.edu/abs/2016ApJ...830..115E}
}

@ARTICLE{fernandez25,
       author = {{Fern\textbackslash'andez-Alvar}, Emma and {Ruiz-Lara}, Tom\textbackslash'as and {Gallart}, Carme and {Cassisi}, Santi and {Surot}, Francisco and {Gonz\textbackslash'alez-Koda}, Yllari K. and {Callingham}, Thomas M. and {Queiroz}, Anna B. and {Battaglia}, Giuseppina and {Thomas}, Guillaume and {Chiappini}, Cristina and {Hill}, Vanessa and {Dodd}, Emma and {Helmi}, Amina and {Aznar-Menargues}, Guillem and {de la Cueva}, Alejandro and {Mirabla}, David and {Quintana-Ansaldo}, M\textbackslash'onica and {Rivero}, Alicia},
        title = "{Chronology of our Galaxy from Gaia colour-magnitude diagram fitting (ChronoGal) II. Unveiling the formation and evolution of the kinematically selected Thick and Thin Discs}",
      journal = {A\&A, submitted},
         year = 2025,
        month = mar,
          eid = {arXiv:2503.19536},
        pages = {arXiv:2503.19536},
          doi = {10.48550/arXiv.2503.19536},
archivePrefix = {arXiv},
       eprint = {2503.19536},
 primaryClass = {astro-ph.GA},
       adsurl = {https://ui.adsabs.harvard.edu/abs/2025arXiv250319536F}
}

@ARTICLE{feltzing19,
       author = {{Feltzing}, Sofia and {Bowers}, J. Bradley and {Agertz}, Oscar},
        title = "{Constraining churning and blurring in the Milky Way using large spectroscopic surveys - an exploratory study}",
      journal = {\mnras},
         year = 2020,
        month = feb,
       volume = {493},
       number = {1},
        pages = {1419-1433},
          doi = {10.1093/mnras/staa340},
archivePrefix = {arXiv},
       eprint = {1907.08011},
 primaryClass = {astro-ph.GA},
       adsurl = {https://ui.adsabs.harvard.edu/abs/2020MNRAS.493.1419F}
}

@ARTICLE{frankel20,
       author = {{Frankel}, Neige and {Sanders}, Jason and {Ting}, Yuan-Sen and {Rix}, Hans-Walter},
        title = "{Keeping It Cool: Much Orbit Migration, yet Little Heating, in the Galactic Disk}",
      journal = {\apj},
         year = 2020,
        month = jun,
       volume = {896},
       number = {1},
          eid = {15},
        pages = {15},
          doi = {10.3847/1538-4357/ab910c},
archivePrefix = {arXiv},
       eprint = {2002.04622},
 primaryClass = {astro-ph.GA},
       adsurl = {https://ui.adsabs.harvard.edu/abs/2020ApJ...896...15F}
}

@ARTICLE{frankel19,
       author = {{Frankel}, Neige and {Sanders}, Jason and {Rix}, Hans-Walter and {Ting}, Yuan-Sen and {Ness}, Melissa},
        title = "{The Inside-out Growth of the Galactic Disk}",
      journal = {\apj},
         year = 2019,
        month = oct,
       volume = {884},
       number = {2},
          eid = {99},
        pages = {99},
          doi = {10.3847/1538-4357/ab4254},
archivePrefix = {arXiv},
       eprint = {1909.07118},
 primaryClass = {astro-ph.GA},
       adsurl = {https://ui.adsabs.harvard.edu/abs/2019ApJ...884...99F}
}

@ARTICLE{frankel18,
   author = {{Frankel}, N. and {Rix}, H.-W. and {Ting}, Y.-S. and {Ness}, M. and 
	{Hogg}, D.~W.},
    title = "{Measuring Radial Orbit Migration in the Galactic Disk}",
  journal = {\apj},
archivePrefix = "arXiv",
   eprint = {1805.09198},
     year = 2018,
    month = oct,
   volume = 865,
      eid = {96},
    pages = {96},
      doi = {10.3847/1538-4357/aadba5},
   adsurl = {http://adsabs.harvard.edu/abs/2018ApJ...865...96F}
}

@ARTICLE{friedli93,
   author = {{Friedli}, D. and {Benz}, W.},
    title = "{Secular evolution of isolated barred galaxies. I - Gravitational
coupling between stellar bars and interstellar medium}",
  journal = {\aap},
     year = 1993,
    month = feb,
   volume = 268,
    pages = {65-85},
   adsurl = {http://adsabs.harvard.edu/abs/1993A%26A...268...65F}
}

@ARTICLE{gonzalez14,
       author = {{Gonz{\'a}lez Delgado}, R.~M. and {P{\'e}rez}, E. and {Cid Fernandes}, R. and {Garc{\'\i}a-Benito}, R. and {de Amorim}, A.~L. and {S{\'a}nchez}, S.~F. and {Husemann}, B. and {Cortijo-Ferrero}, C. and {L{\'o}pez Fern{\'a}ndez}, R. and {S{\'a}nchez-Bl{\'a}zquez}, P. and {Bekeraite}, S. and {Walcher}, C.~J. and {Falc{\'o}n-Barroso}, J. and {Gallazzi}, A. and {van de Ven}, G. and {Alves}, J. and {Bland-Hawthorn}, J. and {Kennicutt}, R.~C. and {Kupko}, D. and {Lyubenova}, M. and {Mast}, D. and {Moll{\'a}}, M. and {Marino}, R.~A. and {Quirrenbach}, A. and {V{\'\i}lchez}, J.~M. and {Wisotzki}, L.},
        title = "{The star formation history of CALIFA galaxies: Radial structures}",
      journal = {\aap},
         year = 2014,
        month = feb,
       volume = {562},
          eid = {A47},
        pages = {A47},
          doi = {10.1051/0004-6361/201322011},
archivePrefix = {arXiv},
       eprint = {1310.5517},
 primaryClass = {astro-ph.CO},
       adsurl = {https://ui.adsabs.harvard.edu/abs/2014A&A...562A..47G}
}

@ARTICLE{geron24,
       author = {{G{\'e}ron}, Tobias and {Smethurst}, R.~J. and {Lintott}, Chris and {Masters}, Karen L. and {Garland}, I.~L. and {Mengistu}, Petra and {O'Ryan}, David and {Simmons}, B.~D.},
        title = "{The Effects of Bar Strength and Kinematics on Galaxy Evolution: Slow Strong Bars Affect Their Hosts the Most}",
      journal = {\apj},
         year = 2024,
        month = oct,
       volume = {973},
       number = {2},
          eid = {129},
        pages = {129},
          doi = {10.3847/1538-4357/ad66b7},
archivePrefix = {arXiv},
       eprint = {2405.05960},
 primaryClass = {astro-ph.GA},
       adsurl = {https://ui.adsabs.harvard.edu/abs/2024ApJ...973..129G}
}

@ARTICLE{kennicutt98,
       author = {{Kennicutt}, Jr., Robert C.},
        title = "{The Global Schmidt Law in Star-forming Galaxies}",
      journal = {\apj},
         year = 1998,
        month = may,
       volume = {498},
       number = {2},
        pages = {541-552},
          doi = {10.1086/305588},
archivePrefix = {arXiv},
       eprint = {astro-ph/9712213},
 primaryClass = {astro-ph},
       adsurl = {https://ui.adsabs.harvard.edu/abs/1998ApJ...498..541K}
}

\end{document}